\documentclass[]{article}

\usepackage{amsfonts}
\usepackage{amsmath}
\usepackage{amssymb}
\usepackage{graphicx}

\usepackage{quoting}
\usepackage{wrapfig}
\usepackage{url}

\usepackage{chngcntr}
\counterwithout{footnote}{section}
\counterwithout{footnote}{subsection}
\counterwithout{footnote}{subsubsection}

\usepackage{amsthm}

\newtheoremstyle{mystyle}
  {}
  {}
  {\itshape}
  {}
  {\bfseries}
  {.}
  { }
  {}

\theoremstyle{mystyle}

\newtheorem{postulate}{Postulate}
\newtheorem{mydef}{Definition}

\newcommand{\CHAIN}[1]{\mathbf{#1}}

\newcommand{\Q}{{\CHAIN{Q}}}
\renewcommand{\P}{{\CHAIN{P}}}
\renewcommand{\S}{{\CHAIN{S}}}
\newcommand{\PI}{{\CHAIN{\Pi}}}

\newcommand{\move}[1]{#1}

\mathchardef\Emptyset="001F   

\newcommand{\dual}{\overline}
\def\JOIN{\mathop\vee}
\def\MEET{\mathop\wedge}

\def\bor{\mathop{\mathord{\lor}\!\!\!\raise4pt\hbox{$\scriptscriptstyle 2$}\,}}
\def\band{\mathop{\mathord{\land}\!\!\!\lower2pt\hbox{$\scriptscriptstyle 2$}\,}}
\def\OR{{\tt OR}}
\def\AND{{\tt AND}}

\def\mydot{ \mathrel{\raisebox{2pt}{\scalebox{0.5}{\textbullet}}}}

\newcommand{\seq}[2]{\ensuremath{[#1,#2]}}   

\newcommand{\lseq}[3]{\ensuremath{[#1,#2,#3]}}   

\title{Information-Based Physics:\\ An Observer-Centric Foundation}
\author{
        Kevin H. Knuth \\
        Departments of Physics and Informatics\\
        University at Albany (SUNY)\\
        Albany NY 12222, USA
}
\date{\today}

\begin{document}
\maketitle

\begin{abstract}
It is generally believed that physical laws, reflecting an inherent order in the universe, are ordained by nature.  However, in modern physics the observer plays a central role raising questions about how an observer-centric physics can result in laws apparently worthy of a universal nature-centric physics.  Over the last decade, we have found that the consistent apt quantification of algebraic and order-theoretic structures results in calculi that possess constraint equations taking the form of what are often considered to be physical laws.  I review recent derivations of the formal relations among relevant variables central to special relativity, probability theory and quantum mechanics in this context by considering a problem where two observers form consistent descriptions of and make optimal inferences about a free particle that simply influences them. I show that this approach to describing such a particle based only on available information leads to the mathematics of relativistic quantum mechanics as well as a description of a free particle that reproduces many of the basic properties of a fermion.  The result is an approach to foundational physics where laws derive from both consistent descriptions and optimal information-based inferences made by embedded observers.
\end{abstract}

\section{Introduction}
It is generally believed that physical laws reflect an inherent order in the universe.  These laws, thought to apply everywhere, are typically considered to have been ordained by nature.  In this sense they are universal and nature-centric.  However, in the last century, modern physics has placed the observer in a central role resulting in an observer-based physics, which with a potential for subjectivity as well as quantum contextuality, poses conceptual difficulties for reconciliation with a nature-based physics. This raises questions as to precisely how an observer-based physics could give rise to consistent universal laws of nature as well as what role information plays in physics \cite{Wiener:1936observer}\cite{Brillouin:1956science}\cite{Jaynes:InfoTheory}\cite{Baierlein:1971}\cite{Tribus+McIrvine:1971sciam}\cite{zurek:1990}\cite{brassard:2005information}\cite{Pawlowski+etal:2009information-causality}\cite{Schumacher+Westmoreland:2010}\cite{Knuth:infophysics}\cite{Goyal:2012information}.  Perhaps the potential implication of such questions has never been so clearly and concisely put as in Wheeler's aphorism ``It from Bit'' \cite{wheeler:1990}.

Over the last decade, we have found that the consistent apt quantification of algebraic and order-theoretic structures results in calculi defined by constraint equations taking the form of what are often considered to be physical laws.  These constraint equations typically result as the solution to functional equations \cite{Aczel:FunctEqns}\cite{aczel+dhombres:1989}, which reflect underlying symmetry or order.  Functional equations have been applied in physics since its mathematical inception with Galileo in 1638 who solved the functional equation
\begin{equation}
\frac{x((n+1)t)-x(nt)}{x(nt)-x((n-1)t)} = \frac{2n+1}{2n-1}
\end{equation}
for the function $x(t)$, demonstrating that falling bodies follow a quadratic law \cite{galilei:dialogues}\cite{aczel+dhombres:1989}.  Since then, they have been used in a variety of specialized contexts including for example d'Alembert's study of vibrating strings in 1747 \cite{aczel+dhombres:1989} and Jaynes' derivation of the Carnot reversible efficiency \cite{Jaynes:Carnot}.  Functional equations have also been applied in more general contexts, such as in Cox's derivation of probability theory \cite{Cox:1946}\cite{Cox:1961} and Pfanzagl's `Theory of Measurement' \cite{Pfanzagl:1968}.

Inspired by Cox's derivation of probability theory as a real-valued quantification of a Boolean algebra, it was realized that more general theories consisting of sets of laws might be derivable by consistent apt quantification of other algebraic structures, the symmetries of which result in sets of functional equations under quantification \cite{Knuth:laws}\cite{Knuth:measuring}\cite{Knuth:infophysics}.  I explored this early on by considering the quantification of an algebra of questions \cite{Knuth:Questions}, again inspired by Cox \cite{Cox:1979}, which led to an inquiry calculus that naturally generalized information theory \cite{Knuth:duality} that was later conceptually and technically refined \cite{Knuth:WCCI06}\cite{Knuth:me08}\cite{vanErp-2013}.

Cox's derivation of probability theory also inspired attempts to derive the rules of manipulating quantum amplitudes and relating them to probabilities.  Notable efforts relying on the methodology of consistent apt quantification were made by Tikochinsky \cite{tikochinsky:1988} along with Gull \cite{tikochinsky:2000}, and independently by Caticha \cite{Caticha:1998}, who quantified an algebra of combining experimental setups using complex numbers.  More advanced order-theoretic derivations of probability theory \cite{Knuth:measuring}\cite{Knuth&Skilling:2012} inspired a derivation of the Feynman rules for the manipulation of quantum amplitudes \cite{Feynman:1948}\cite{Feynman&Hibbs} based on consistent quantification of measurement sequences using real-valued pairs \cite{GKS:PRA}\cite{GK:Symmetry}. 

We later extended these efforts to more general order-theoretic structures, such as causally-ordered sets.  This was performed by relating elements of a partially-ordered set to a distinguished chain representing an observer.  In special cases where two observer chains consistently quantify the lengths of each other's intervals, consistent apt quantification results in the mathematics of flat Minkowski spacetime and Lorentz transformations \cite{Knuth+Bahreyni:EventPhysics}\cite{Bahreyni:Thesis}.

While our past efforts focused on quantifying more general symmetries and order-theoretic structures as a route to understanding physical laws, physics as applied to the real world ultimately must describe \emph{something}, and that something can only be modeled by making basic assumptions about its nature.  My most recent work on causally-ordered sets suggests a model for a free particle \cite{Knuth:fermions}, based on a primitive concept of causality where particles simply influence one another \cite{Knuth:FQXI2013} in a discrete fashion similar in spirit to the direct particle-particle interaction models of Wheeler and Feynman \cite{Wheeler+Feynman:1945}\cite{Wheeler+Feynman:1949}.  In this paper, I demonstrate the methodology of consistent apt quantification by applying it to this purposefully simple model of a free particle.  In doing so, I will summarize our previous work in deriving special relativity, probability theory and quantum mechanics and show that application of these techniques to an influence-based free particle model results in important aspects of fermion physics.  This is accomplished by considering a free particle that influences two observers in a discrete fashion.  I show that equipped only with information about the fact that they were influenced, these observers can only consistently describe the particle using relevant variables analogous to observable fermion properties and the mathematics of relativistic quantum mechanics.  The resulting problem formulation is essentially the Feynman checkerboard model of the electron \cite{Knuth:fermions}\cite{Knuth:FQXI2013}, which is known to lead to the Dirac equation in 1+1 dimensions \cite{Feynman&Hibbs}\cite{gersch1981feynman}\cite{Jacobson+Schulman:1984}\cite{ord1993dirac}\cite{Kauffman:1996:DiscreteDirac}\cite{ord2002feynman}\cite{Earle:DiracMaster2011}.

The remainder of this paper is organized into four main sections: \emph{Model}, \emph{Description} and \emph{Inference}, followed by a short \emph{Discussion}.  Section \ref{sec:Model}, titled \emph{Model}, introduces and motivates the influence model of a free particle and explains how it results in a partially-ordered set.  Section \ref{sec:Description}, titled \emph{Description}, focuses on a consistent observer-based description of the behavior of a free particle.  In this section, I review how consistent apt quantification of partially-ordered sets by embedded observers leads to a concept of emergent spacetime, and demonstrate how the application of consistent quantification to the influence-based free particle model results in several fundamental fermion properties.  Section \ref{sec:Inference}, titled \emph{Inference}, considers optimal inferences made by embedded observers about the free particle's behavior based on measurement sequences.  This includes a review of a derivation of probability theory as a consistent quantification of relationships between logical statements.  This is followed by a review of the consistent quantification of measurement sequences, which together with probability theory results in the Feynman rules for manipulating quantum amplitudes.  These results are then applied to the influence-based model of a free particle to demonstrate how this leads to the Feynman checkerboard model of the Dirac equation.  The result is an approach to foundational physics where laws derive from both consistent descriptions and optimal information-based inferences made by embedded observers.

\section{Model} \label{sec:Model}

\subsection{Influence}  \label{sec:Influence}
We know about particles, such as electrons, because they influence our measurement apparatus.  We endow them with properties, such as mass, momentum and spin, and identify the laws of physics that describe the mathematical inter-relationships among these relevant variables.  In the course of this endeavor, we have become so familiar with these variables that we often view them as fundamental.  However, there are difficulties.  For instance, quantum mechanics teaches us that an electron cannot simultaneously possess both a definite position and a definite momentum, which is referred to as complementarity \cite{Einstein-EPR-1935}\cite{Bohr:1935}.  The concept of quantum contextuality takes this even further since the measurement result of an observable of a system is dependent on the precise way in which it is measured \cite{Kochen+Specker:1967}.  These concepts run counter to the notion that a property is something that is possessed.  Is there perhaps a different way to think about these particles and their properties?

Imagine that electrons are pink and fuzzy, but that their pinkness and fuzziness does not affect how they influence any measurement apparatus.  In this thought experiment, the fact that electrons are pink and fuzzy is inaccessible and thus unknowable.  The result is that the only accessible ``properties'' of an electron are those that affect how an electron influences others.  One could adopt an operationalist definition where such properties are described by patterns or qualities of an electron's influence.  That is, instead of focusing on what an electron \emph{is} and what properties it possesses, perhaps it would be better to simply focus on what an electron \emph{does}.

\begin{figure}[t]
  \begin{center}
  \includegraphics[height=0.20\textheight]{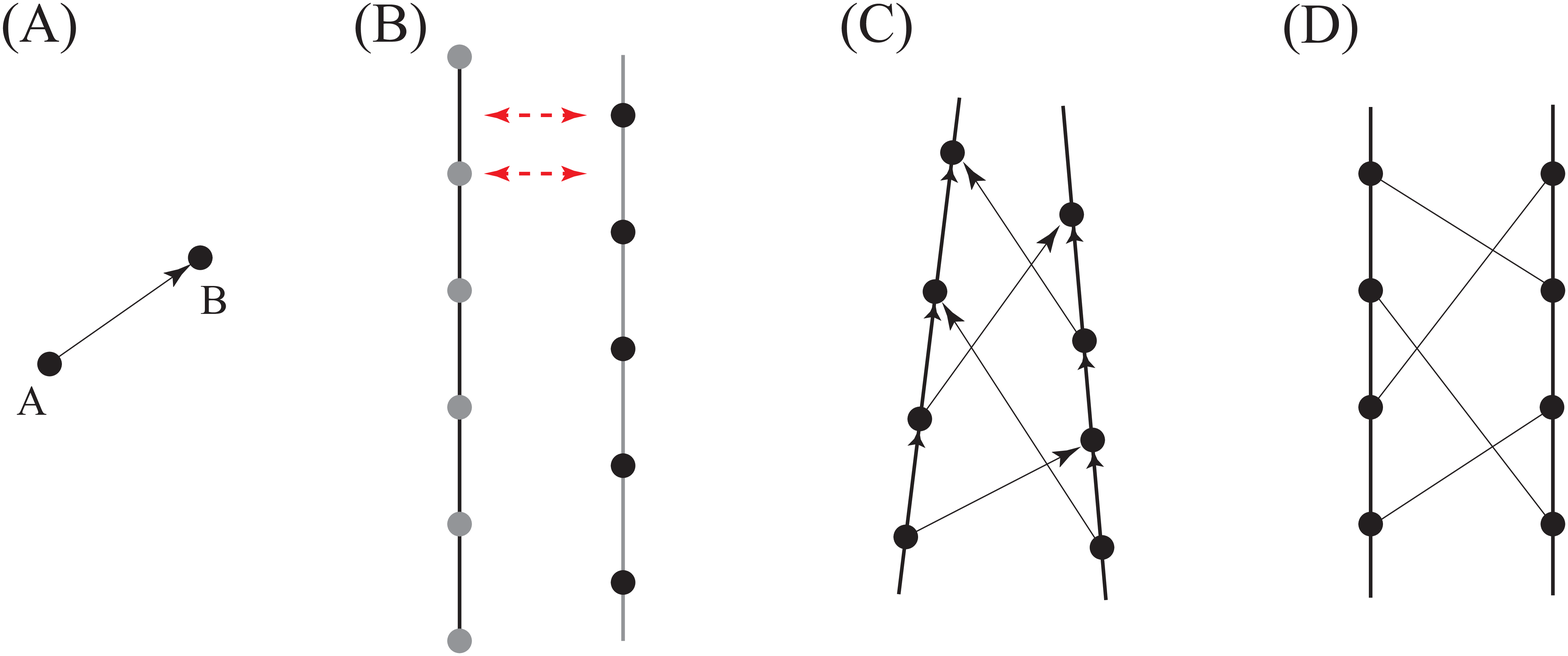}
  \end{center}
  \caption{(A) An illustration of Postulate 1 where one particle influences another. The act of the first fermion influencing is represented by event A and the response of the second fermion to such influence is represented by event B.  Here we illustrate the ordering `A influences B'.  The dual ordering given by `B is influenced by A' would be illustrated by simply reversing the arrow.  (B) An illustration of Postulates 2 and 3. (Left chain) States are affected by events so that events (black lines) couple states (gray circles) allowing states to be totally ordered forming a chain where the selected direction of ordering (up or down) is arbitrary.  (Right chain) Dually, events are affected by states so that states (gray lines) couple events (black circles) allowing events to be totally ordered forming a chain again with an arbitrary ordering direction.  The red arrows highlight the duality graphically as lines are mapped to circles and vice versa.  A particle can be represented with either model.  We will choose to use a chain of events with events represented by circles. (C) This directed acyclic graph illustrates two particles influencing one another.  The two ordering relations are considered together with ordering selected to ensure that the directed graph is acyclic.  Two different arrow shapes highlight the two different ordering relations.  Reversing all the arrows results in the dual ordering.  (D) This is a Hasse diagram of the poset representing the two particles in (C) with the events ordered causally.  The arrows are omitted with the understanding that the ordering relation is directed upward.  Flipping the poset upside-down reverses the causal ordering relation.  Note that the only important feature is the connectivity.  Lengths, distances, and orientation of elements in the illustration are irrelevant.}
  \label{fig:postulates}
\end{figure}

\subsection{Direct Particle-Particle Influence Model}
I introduce a simple model of influence, which focuses on what it means for something to influence something else.  This is meant to be a more primitive concept than that of interaction or force; although, as I will describe later, they are not unrelated.
\begin{postulate} \label{post:influence}
Particles can influence one another in a pair-wise fashion, such that one particle influences and the other is influenced.
\end{postulate}
This postulate assumes that particles can influence one another in such a way that is not symmetric.  Both the specific act of one particle influencing another and the specific response of the other particle to such influence can be identified and distinguished.  This leads to the definition of an event.
\begin{mydef}[Event]
A given instance of influence defines two events, one associated with the act of influencing and the other associated with the response to such influence.
\end{mydef}
Every instance of influence can be described by two events, each associated with one of the two participating particles.  These events can be ordered based on their relationship to one another as either: `A influences B' or `B is influenced by A' (Figure \ref{fig:postulates}A).  Note that events are not assumed to take place in any space or time.

Next I must make some assumption about the events experienced by a particle.  There are a few ways one could do this.  The most direct way would be to assume that the events experienced by a particle are totally ordered in some sense.  However, such an assumption lacks motivation, suggests that other partial orders might be reasonable to consider, and creates the impression that I am assuming a concept of time at the outset where the events follow some kind of temporal sequence.  Instead one could focus on the concept of change by postulating that a particle carries with it some internal state that is changed by either an act of influence or a response to such influence.  Events would then represent change points in the particle state.  However, the difficulty here again is the fact that the concept of change implies that there is some natural ordering: before and after.  It may be beneficial to attempt to back up even further and simply assume that a particle has associated with it some internal state that couples pairs of events, so that \emph{two events bound a state}.  Dually, events can be conceived to couple states so that \emph{two states bound an event}.\footnote{It may be that events affect states and states affect events, but since we have no access to internal states, we will focus only on the relationship between pairs of events coupled by a single state.}  I assume transitivity and insist that no pair of events or states can act as boundaries in two different ways, which rules out cyclic behavior.  The result is then that both states and events can be totally ordered, forming a chain (Figure \ref{fig:postulates}B).
\begin{postulate}  \label{post:state-changed-by-events}
A particle has associated with it an internal state such that each influence event uniquely relates one particle state to another, such that each influence event is bounded by two particle states.
\end{postulate}
\begin{postulate} \label{post:state-couples-events}
Each particle state uniquely relates one influence event to another, such that each particle state is bounded by two influence events.
\end{postulate}
\begin{postulate} \label{post:transitivity}
The roles that two events (or states) play in forming the boundaries of a state (or event) can be distinguished from one another so that the two boundary elements can be ordered.  This ordering is assumed to be transitive, such that no pair of boundary elements can be ordered in both ways.
\end{postulate}
These postulates imply that the set of events that a particle experiences can be totally ordered in a non-cyclic fashion thus forming a chain of events coupled by states, which can be used to represent a particle in this model.  It will be shown later that this totally ordered set of events gives rise to a concept of time, which is more elaborate.


At this point, there are two ordering relations that can be used to order events.  The act of influence allows one to order pairs of events experienced by different particles; whereas the events experienced by a given particle can be totally ordered.  This is illustrated in Figure \ref{fig:postulates}C, which shows two particles influencing one another.  Note that the two types of ordering are illustrated using different arrows, and that the direction of the ordering was selected so that there are no loops, which would lead to contradictions.  The result is that together the two ordering relations form a directed acyclic graph (DAG), which can be represented as a partially ordered set, or poset, of events along with a binary ordering relation (Figure \ref{fig:postulates}D), which in this case is the union of the two ordering relations set up in Postulates \ref{post:influence}-\ref{post:transitivity}.  The next definition will consider both of these ordering relations to be examples of causal ordering.
\begin{mydef}[Causal Ordering] \label{def:causality}
Influence induced order and ordering along a particle chain are members of an equivalence class referred to as causal ordering, denoted by $<$, where the relative directions of the two ordering relations are defined to prohibit contradictory cyclic behavior.
\end{mydef}
Note that the direction of causal ordering can be defined in two ways so that the dual order can be obtained by reversing all the arrows.  This is the origin of time-reversal symmetry, though in order to maintain invariance other transformations would also need to be applied.
Last, I postulate transitivity of causal ordering so that these effects can be chained up.
\begin{postulate} \label{post:causal-transitivity}
The causal ordering relation is assumed to be transitive so that if $X < Y$ and $Y < Z$ then $X < Z$, where $X$, $Y$, and $Z$ may or may not be elements of a single particle chain.
\end{postulate}

The result is a purposefully simplistic model of particles that influence one another in a discrete fashion.  Together, they give rise to a poset of events (Figure \ref{fig:postulates}D) forming a causally-ordered set, which has been referred to as a causet \cite{Bombelli-etal-causal-set:1987}\cite{Bombelli-etal-origin-lorentz:1989}\cite{Sorkin:2003}\cite{Sorkin:2006}.  However, it should be emphasized that it is not assumed that these events take place in any space or time.  Instead, the goal is to develop all aspects of the theory from the bottom up.  I will review each step of the process in sufficient detail to illustrate how quantification of order-theoretic structures enables one to derive laws that reproduce a surprising amount of physics.

\section{Description} \label{sec:Description}
The direct particle-particle influence model represents a particle as a chain of events embedded within a causally-ordered poset obeying specific connectivity rules arising from the fact that the process of influence relates pairs of chains in a discrete fashion.  At the most fundamental level, an observer obtains information about the universe through interactions that influence particles in his or her sensory organs or measurement apparatus.  I model an ideal observer as one who has access to a single particle chain and can keep track of events where that particle influences others or is influenced by others. This is referred to as an \emph{\textbf{observer chain}}.

In this section I review previous work demonstrating how a subset of the poset of events is aptly and consistently quantified by an observer chain.  The results, which amount to the mathematics of flat spacetime, will be applied to the free particle model to obtain a \textbf{consistent observer-based description of particle behavior}.  The interested reader is encouraged to consult the more detailed treatise on which this summary is based \cite{Knuth+Bahreyni:EventPhysics}\cite{Bahreyni:Thesis}.

\subsection{Quantification of Partially Ordered Sets: Emergent Spacetime} \label{sec:spacetime}
In this section, I obtain a consistent observer-based (chain-based) quantification of a poset.  I will consider posets that contain at least one chain and enjoy a more general connectivity than that specified by the postulates for influence above.

\subsubsection{Quantifying a chain} \label{sec:chain-quantification}
I begin by considering the consistent quantification of a finite chain of elements.  I will work through this in some detail to demonstrate how, in this more simple case, order and symmetries are used to obtain a consistent means of quantification.

An element of a chain $\P$ is \textbf{\emph{quantified}} by defining a functional $v_{\P}$, called an \textbf{\emph{isotonic valuation}}, that assigns a real number to each element $p$ of the chain where $v_{\P} : p \in \P \rightarrow \mathbb{R}$, such that for every $p_x, p_y \in \P$ with the relationship $p_x < p_y$ we have that $v_{\P}(p_x) \leq v_{\P}(p_y)$.\footnote{The case where $v_{\P}(p_x) = v_{\P}(p_y)$ is intentional as it allows for coarse graining.}  The valuation takes the $N$ elements of the finite chain $\P$, $p_1, p_2, \ldots, p_N$, to a sequence of real numbers $v_{\P}(p_1) \leq v_{\P}(p_2) \leq \ldots \leq v_{\P}(p_N)$.  To simplify the notation, I will overload the symbol that labels an element, such as $p$, by using it to also represent the valuation $v_{\P}(p)$ assigned to that element where it will be understood from context whether the symbol $p$ refers to the element or its real-valued valuation.

Since a finite chain is isomorphic to a subset of the natural numbers ordered by the usual `less than or equal to', no generality is lost by assigning valuations that amount to counting.  Later, I will show that, in general, real numbers are required.  This can be handled by choosing a real number $\mu$ to represent a basic unit of measure.  Therefore, an element $p_k$ of the chain $\P$ can, without loss of generality, be assigned a valuation
\begin{equation}
v_{\P}(p_k) \equiv p_k := k\mu
\end{equation}
where $k \in \mathbb{Z}$ is an index that counts the elements and $\mu$ is an arbitrary real number.  To be clear, note that the symbol $p_k$ on the left-hand side of the equation refers to the chain element and the symbol $p_k$ on the right-hand side of the equation refers to its real-valued valuation.

A \textbf{\emph{closed interval}} between two elements $p_i$ and $p_j$ of a chain $\P$, denoted $[p_i, p_j]_\P$, is defined as the set of all elements $x \in P$ such that $p_i \leq x \leq p_j$.  That is,
\begin{equation}
[p_i, p_j]_\P := \{p_i, p_{i+1}, \ldots, p_j\}.
\end{equation}
One can quantify closed intervals similarly by defining an isotonic valuation $\phi$, which is a functional that takes a closed interval $I_\P$ to a real number $\phi(I_\P)$ such that $\phi(J_\P) \leq \phi(I_\P)$ if $J_\P \subseteq I_\P$.
Since a closed interval is a set, the valuation must be a function of the cardinality of that set.

General rules apply to specific cases. Thus one can constrain the form of a general rule by considering an appropriate special case.  Here I consider joining two intervals that share a single element, which can be written as a set union
\begin{eqnarray}
[p_i,p_j]_{\P} \cup [p_j,p_k]_{\P} & := & \{p_i, p_{i+1}, \ldots, p_j\} \cup \{p_j, p_{j+1}, \ldots, p_k\} \\
 & = & \{p_i, p_{i+1}, \ldots, p_k\} \\
 & = & [p_i, p_k]_{\P}.
\end{eqnarray}
By writing $I = [p_i, p_k]_{\P}$, $J = [p_i, p_j]_{\P}$, $K = [p_j, p_k]_{\P}$, the expression above can be written as $I = J \cup K$.
I impose a consistency requirement that the valuation quantifying the resulting closed interval $I$ must be some function of the valuations quantifying the two intervals, $J$ and $K$, that were joined.  If this were not the case, then the valuation would not encode the relationship between an interval and any of its sub-intervals.  This relation can be expressed as
\begin{equation} \label{eq:lengths-oplus}
\phi(I) = \phi(J) \oplus \phi(K)
\end{equation}
where $\oplus$ represents an unknown function to be determined.

I can consider joining a third closed interval, $L = [p_k, p_m]_{\P}$ sharing only the element $p_k$ with $K = [p_j,p_k]_{\P}$, and note that the joining process obeys associativity so that the order in which closed intervals are joined does not matter
\begin{equation}
(J \cup K) \cup L = J \cup (K \cup L).
\end{equation}
By applying (\ref{eq:lengths-oplus}), the valuations reveal that the function $\oplus$ must also be associative
\begin{equation}
(\phi(J) \oplus \phi(K)) \oplus \phi(L) = \phi(J) \oplus (\phi(K) \oplus \phi(L)),
\end{equation}
which is made more apparent by writing $\alpha = \phi(J)$, $\beta = \phi(K)$ and $\gamma = \phi(L)$ so that
\begin{equation} \label{eq:oplus-associativity}
(\alpha \oplus \beta) \oplus \gamma = \alpha \oplus (\beta \oplus \gamma).
\end{equation}
Equation (\ref{eq:oplus-associativity}) is a functional equation for the function $\oplus$, which is known as the \textbf{\emph{associativity equation}} \cite{Aczel:FunctEqns}\cite{aczel+dhombres:1989}\cite{Knuth&Skilling:2012}.  Its general solution can be written as
\begin{equation} \label{eq:associativity-soln-relativity}
\alpha \oplus \beta = f^{-1}(f(\alpha) + f(\beta))
\end{equation}
where $f$ is an arbitrary invertible function \cite{Aczel:FunctEqns}\cite{aczel+dhombres:1989}\cite{Knuth&Skilling:2012}. This means that there exists some invertible function $f$, which allows one to perform a regraduation of the valuation $\phi$ to a more convenient additive valuation $d = f \circ \phi$ so that $d(I) = f(\phi(I))$ and
\begin{equation} \label{eq:additivity-of-closed-intervals}
d(I) = d(J) + d(K),
\end{equation}
whenever $I = J \cup K$ and the cardinality of $J \cap K$ is one.  I refer to this valuation $d$ of closed intervals along a chain as the \textbf{\emph{length}} of the interval.

A further requirement is that the length of a closed interval is some function of the valuations assigned to the endpoint elements defining the closed interval
\begin{equation} \label{eq:length-and-endpoints}
d([a,b]_{\P}) = s(a,b),
\end{equation}
where the function $s$ is to be determined.
Applying (\ref{eq:additivity-of-closed-intervals}) and (\ref{eq:length-and-endpoints}) to $[a,c]_{\P} = [a,b]_{\P} \cup [b,c]_{\P}$ results in a functional equation for the function $s$
\begin{equation}
s(a,c) = s(a,b) + s(b,c).
\end{equation}
Since $s(a,c)$ is not a function of $b$, the general solution can be written as
\begin{equation} \label{eq:closed-interval-endpoint-difference}
s(a,b) = g(b)-g(a)
\end{equation}
where $g$ is an arbitrary function.
Last, since closed intervals are just sets, the length must be some function $h$ of the cardinality of the set so that $d([p_j, p_k]_{\P}) = h(k-j)$.  Applying (\ref{eq:closed-interval-endpoint-difference}) gives
\begin{equation}
g(k\mu) - g(j\mu) = h(k-j)
\end{equation}
so that $g$ must be linear.
As a result, up to arbitrary scale which amounts to selecting units, the length of the interval is found by taking the difference of the valuations of its endpoints
\begin{equation}
d([p_j, p_k]_\P) = p_k-p_j \equiv (k-j)\mu.
\end{equation}

In this context, since we are familiar with the solution, such a result may seem trivial.  However, this derivation reveals precisely what freedom one has in quantifying lengths of intervals along a chain.  That is, we have the choice of choosing units of measure, which amounts to linear rescaling.  In addition, combinations of length must obey some invertible transformation of addition.  Later, we will use the associativity equation several times to obtain more profound results in probability theory and quantum mechanics.

\subsubsection{Chain projection}
\begin{figure}[t]
  \begin{center}
  \includegraphics[height=0.27\textheight]{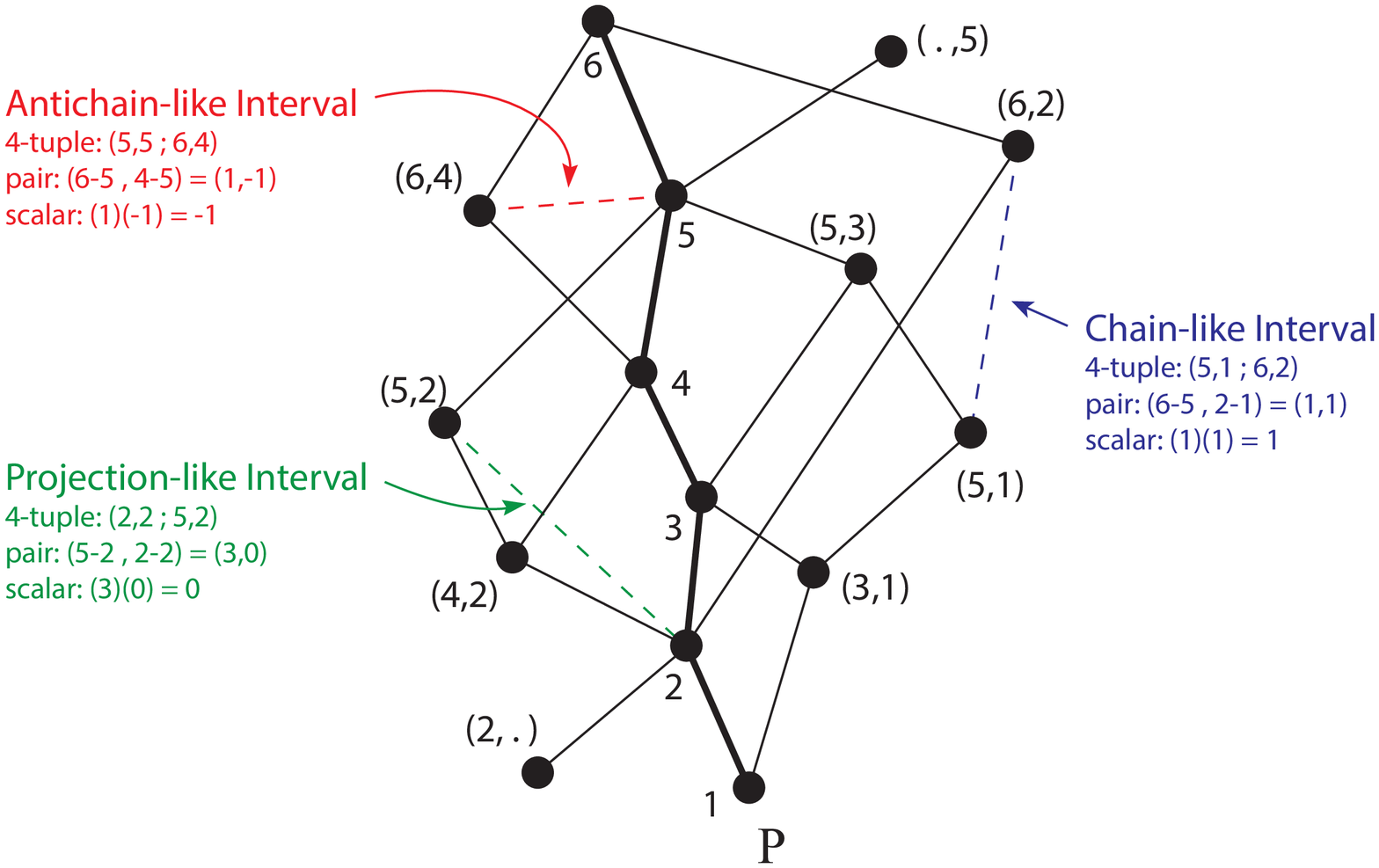}
  \end{center}
  \caption{This figure illustrates the quantification of a poset by an embedded chain $\P$.  Elements comprising the chain are quantified, up to scale, by counting numbers.  Poset elements are potentially quantified by pairs of numbers based on the valuations assigned to the forward projection and backward projection of the element onto the chain.  The result is a chain-based coordinate system that quantifies some subset of the poset.  Not illustrated is the fact that elements on the chain are quantified by a symmetric pair, since an element of a chain projects onto itself.  Note that not all projections onto a given chain exist.  In such cases a non-existent projection is denoted by using a dot.  Intervals are defined by pairs of poset elements (see three specific examples illustrated with colored dashed lines discussed in the text).
  }
  \label{fig:poset-quantification}
\end{figure}

I now examine the relationship between a poset element $x$ and an observer chain $\P$.  First, consider the case where $x$ is included by (causally influences) some element of the chain $\P$.  Since $\P$ is a finite chain, there exists a unique minimum element, $p_x$, that includes (is causally influenced by) $x$.  This allows one to define a functional called the forward projection operator that takes a poset element $x$ from the domain of all elements included by any element of the chain to the unique minimum element on the chain that includes $x$.  I denote the forward projection operator with the same letter that labels the chain, so that the equation $p_x = Px$ indicates that the \emph{\textbf{forward projection}} $P$ of the element $x$ onto the chain $\P$ is the element $p_x \in \P$.
Dually, I can define the \emph{\textbf{backward projection}} to be a functional that takes a poset element in the domain consisting of the set of all elements that include (are causally influenced by) any element of the chain to the greatest element of the chain that is included by it.  Both the operator and the resulting chain element are denoted with a bar, so that one can write $\bar{p_x} = \bar{P}x$,
where $\bar{p_x}$ is the greatest element of the chain $\P$ that is included by (causally influences) $x$.

Since the projections, if they exist, of an element onto a chain are unique, they provide a unique quantification of the poset element in terms of its relationship to the chain.  The result is a chain-based coordinate system that quantifies some subset of the entire poset.  For example, a poset element $x$ that both forward and backward projects to the chain $\P$ is uniquely quantified by the pair of numbers $(v_P(Px), v_P(\bar{P}x)) = (v_P(p_x), v_P(\bar{p}_x)) \equiv (p_x, \bar{p}_x)$, where in the last step I used the fact that the symbol $p_x$ is overloaded so that the label for the element is also used to represent its associated valuation.  Such a quantification of a poset with respect to the chain $\P$ is illustrated in Figure \ref{fig:poset-quantification}.  Note that elements belonging to the observer chain itself are quantified with a symmetric pair (not illustrated) since any element of the chain both forward projects and backward projects to itself.  In addition, quantification with respect to a different chain would change the coordinates of the poset elements.

\subsubsection{Intervals}
The concept of a closed interval along a chain can be extended to the poset as a whole by defining a generalized interval, or \emph{\textbf{interval}} for short, denoted $[a,b]$, as an ordered pair of \emph{any} two poset elements, each of which are referred to as \emph{\textbf{endpoints}}.

In discussing consistent quantification, I consider only the case where both elements comprising the endpoints of an interval both forward and backward project onto the observer chain.  The result is that the interval $[a,b]$ maps to two closed intervals on the chain: $[p_a, p_b]_{\P}$ and $[\bar{p_a}, \bar{p_b}]_{\P}$.  This leads to a 4-tuple quantification of the interval given by the valuations assigned to the four projections of the endpoints onto the chain: $(p_a, \bar{p_a} ; p_b, \bar{p_b})$, where the semicolon is used to separate the two pairs.  Furthermore, each of the closed intervals onto which the interval projects has an associated length.  These lengths provide a natural pairwise quantification of the interval with respect to the observer chain, called the \emph{\textbf{interval pair}}, given by: $(p_b - p_a, \bar{p_b}-\bar{p_a})$.  By writing the lengths as $\Delta p = p_b-p_a$ and $\Delta \bar{p} = \bar{p_b} - \bar{p_a}$, we can write the quantifying interval pair as $(\Delta p, \Delta \bar{p})$.  In the next subsection, I consider a consistent scalar quantification of the interval.

\subsubsection{Interval Scalar}
\begin{figure}[t]
  \begin{center}
  \includegraphics[height=0.27\textheight]{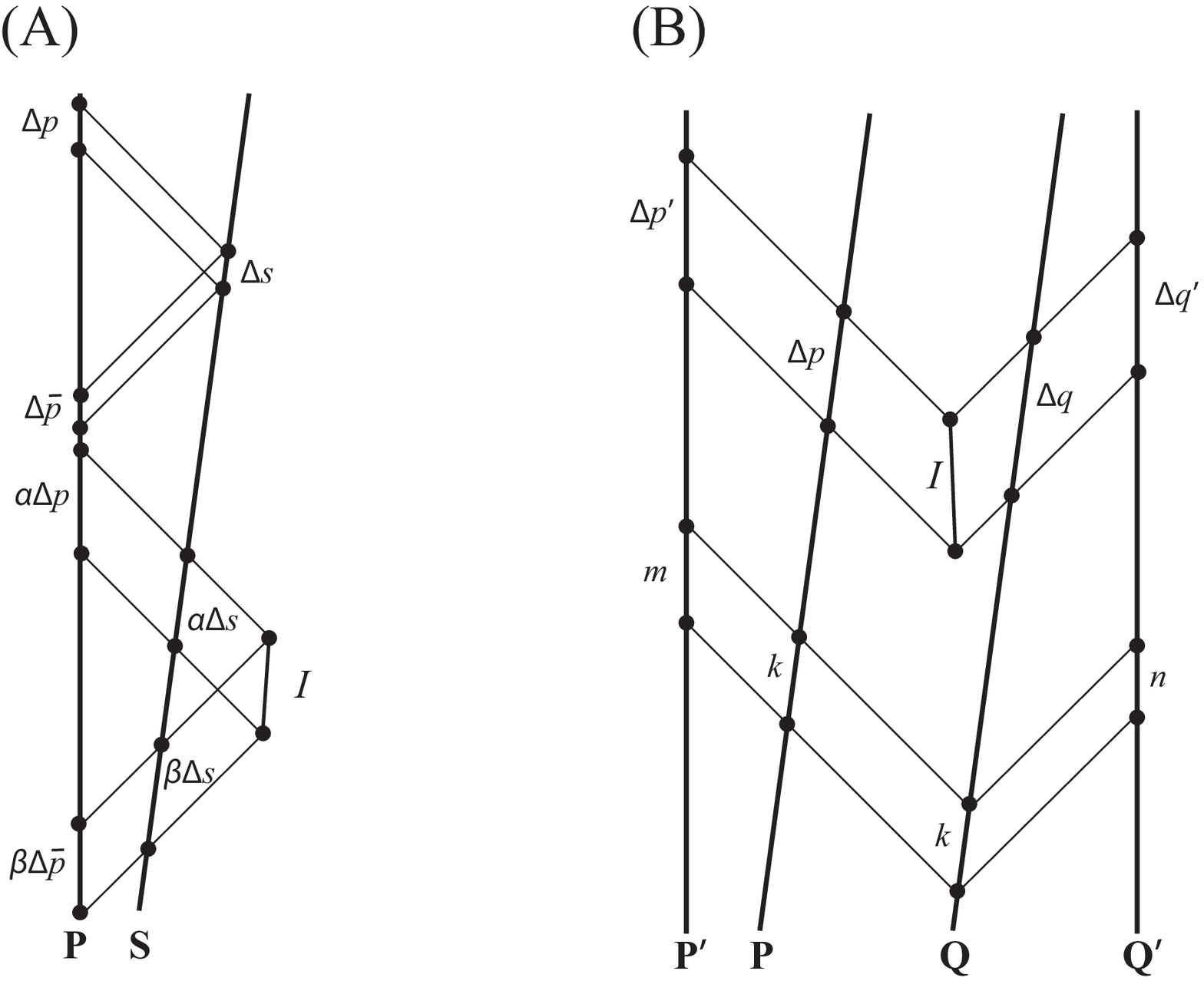}
  \end{center}
  \caption{(A) The two chains $\P$ and $\S$ are related by constant projection where every closed interval of length $\Delta s$ on $\S$ forward projects to a closed interval of length $\Delta p$ on $\P$ and backward projects to a closed interval of length $\Delta \bar{p}$.  Constant projection implies linearity so that the chains $\P$ and $\S$ are said to be linearly-related.  In this situation, consistent quantification requires that there is a functional relationship between quantification by one chain and quantification by the other, which is shown to be a Lorentz transformation. (B) An illustration of the pair transform between the pair of coordinated chains PQ and the linearly-related pair of coordinated chains $P'Q'$ where each closed interval of length $\Delta p = \Delta q = k = \sqrt{mn}$ projects to closed intervals of lengths $\Delta p' = m$ and $\Delta q' = n$.}
  \label{fig:homogeneity}
\end{figure}

Here I aim to constrain the general rule by considering the special case of \emph{\textbf{constant projection}} where two chains $\P$ and $\S$ project to one another in a constant fashion, so that every closed interval of length $\Delta s$ on $\S$ forward projects to a closed interval of length $\Delta p$ on $\P$ and backward projects to a closed interval of length $\Delta \bar{p}$ (Figure \ref{fig:homogeneity}A). Constant projection implies linearity so that an interval $I$ that projects onto the chain $\S$ resulting in the quantifying pair $(\alpha \Delta s, \beta \Delta s)$ with respect to $\S$, projects onto $\P$ resulting in the quantifying pair $(\alpha \Delta p, \beta \Delta \bar{p})$ with respect to $\P$. For this reason, the chains $\P$ and $\S$ are said to be \emph{\textbf{linearly-related}}.

Since a closed interval along a chain is a special case of an interval, I require that its scalar quantification be consistent with its length.  The result is that the scalar measure must be some symmetric function $\sigma$ of the interval pair such that
\begin{equation} \label{eq:scalar-sigma}
\Delta s = \sigma(\Delta p, \Delta \bar{p})
\end{equation}
where $\Delta s$ is the scalar measure and $\sigma$ is an unknown function to be determined.  I have the freedom to rescale the lengths of the closed intervals by a positive real number $c$ so that
\begin{align} \label{eq:functional-equation-for-sigma}
c \Delta s &= \sigma(c \Delta p, c \Delta \dual{p}) \\
&= c \sigma(\Delta p, \Delta \dual{p}), \nonumber
\end{align}
where the second line comes from substituting (\ref{eq:scalar-sigma}) into $\Delta s$ on the first line of (\ref{eq:functional-equation-for-sigma}).  Equating the right-hand sides results in a functional equation for $\sigma$ known as the \emph{\textbf{homogeneity equation}}
\begin{equation}
F(cx, cy) = c^k F(x,y)
\end{equation}
with $k=1$.  Its general solution can be written as \cite{Aczel:FunctEqns}
\begin{equation}
F(x,y) =
\begin{cases}
\sqrt{xy} ~ h(\frac{x}{y}) & \mbox{if } xy \neq 0  \\
ax & \mbox{if } x \neq 0, y = 0 \\
by & \mbox{if } y \neq 0, x = 0 \\
d & \mbox{if } y = 0, x = 0
\end{cases}
\end{equation}
where $h(\frac{x}{y}) = h(\frac{y}{x})$ and $a=b$ since $\sigma$, and hence $F$, is symmetric in its arguments.  When $\Delta s = 0$ we have that $\Delta p = \Delta \overline{p} = 0$ so that $\sigma(0,0) = 0$ giving $d = 0$.  Additional special cases \cite{Knuth+Bahreyni:EventPhysics} enable one to show that $h(\frac{x}{y}) = 1$ for all $x$ and $y$, and that the solution is simply
\begin{equation}
\Delta s = \sqrt{\Delta p \Delta \bar{p}}.
\end{equation}

However, in general, intervals may have projections with negative lengths.  It is possible to define orthogonal subspaces, and associativity in joining orthogonal intervals results in additivity of the scalar interval \cite{Knuth+Bahreyni:EventPhysics}.  The result is that it can be shown that in general the scalar quantification of an interval, referred to as the \emph{\textbf{interval scalar}}, is given by
\begin{equation}
{\Delta s}^2 = \Delta p \Delta \bar{p},
\end{equation}
where ${\Delta s}^2$ is a single symbol and does not denote the square of any quantity, except the length of a closed interval in special cases.
The interval scalar can be used to identify three classes of intervals, which are invariant as one transforms from one observer chain to another.  An interval is \emph{\textbf{chain-like}} if its interval scalar is positive, \emph{\textbf{antichain-like}} if its interval scalar is negative, and \emph{\textbf{projection-like}} if its interval scalar is zero.  Examples of such intervals, depicted using dashed lines, are illustrated in Figure \ref{fig:poset-quantification}.

I now return to consider the pairwise quantification of the interval $I$ discussed in the beginning of this section, and define a function $L$ that takes the pair quantification of an interval with respect to one chain to a pair quantification with respect to another chain.  Given that a closed interval on $\S$ of length $\Delta s = k = \sqrt{mn}$ has projections $\Delta p = m$ and $\Delta \bar{p} = n$ onto $\P$, the \emph{\textbf{pair transform}} from $\S$ to $\P$ can be written as \cite{Knuth+Bahreyni:EventPhysics}
\begin{eqnarray} \label{eq:pair-transform-one-chain}
(\Delta p, \Delta \bar{p})_{\P} &=& L_{\S \rightarrow \P}(\Delta s, \Delta \bar{s})_{\S} \\ \nonumber
&=& \Big(\Delta s \sqrt{\frac{m}{n}}, \Delta \bar{s} \sqrt{\frac{n}{m}} \Big)_{\P}
\end{eqnarray}
where the product of the pairs remains invariant.  At this point, it is clear that simply counting events is not sufficient since real-valued measures are necessary to accommodate such transforms.

\subsubsection{Length and Distance}
%

\begin{figure}[t]
  \begin{center}
  \includegraphics[height=0.22\textheight]{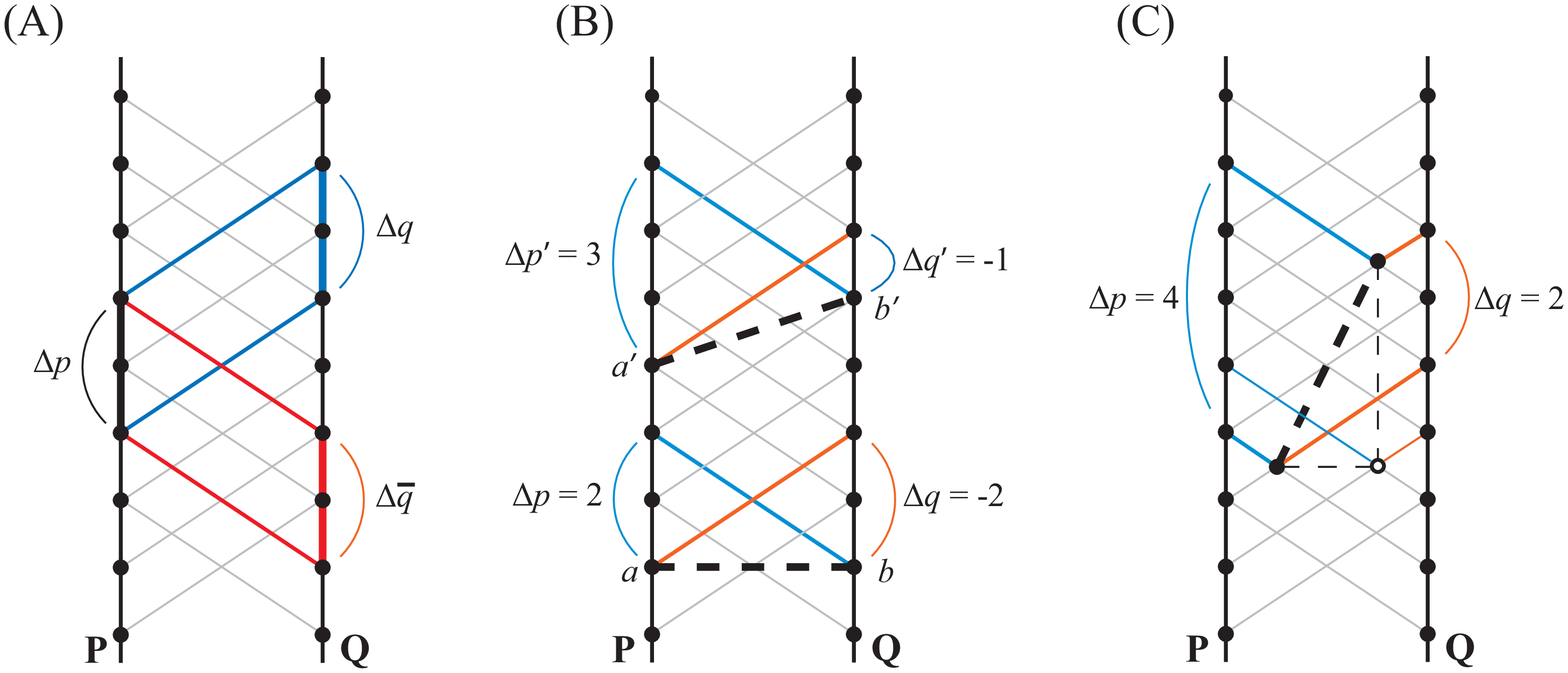}
  \end{center}
  \caption{(A) illustrates the concept of coordination where intervals on one chain project onto intervals of the same length on the other chain and vice versa (not shown). (B) illustrates the distance measure between chains. It is a function of an element on each chain, but cannot depend on which elements were chosen. Intervals $[a,b]$ and $[a',b']$ are shown with distances $D([[P,Q]])$ given by $(\Delta p - \Delta q) / 2 = (2-(-2)) / 2 = 2$ and $(\Delta p' - \Delta q') / 2 = (3-(-1)) / 2 = 2$.  Note also that the interval $[a,b]$ is quantified by the antisymmetric pair $(2,-2)$ and the scalar $(2)(-2) = -4$, which is the reason for the minus sign in the metric (\ref{eq:metric}). (C) illustrates the symmetric-antisymmetric decomposition. An interval quantified by the pair $(4, 2)$ can be decomposed into an interval quantified by the pair $(3,3)$ of length $3$ along the chains and an interval quantified by the pair $(1,-1)$ corresponding to a distance of $1$ between the chains so that $\Delta p \Delta Δq = (4)(2) = (3)(3) + (1)(-1) = 8$.}
  \label{fig:coordination+distance+decomposition}
\end{figure}

In the previous section, I discussed chains that project to one another in a constant fashion.  Here I consider a special case of constant projection called \emph{\textbf{coordination}} where the projections $m=n$ over some finite range, called the coordination range, such that the two chains agree on the length of each other's intervals (Figure \ref{fig:coordination+distance+decomposition}A).  As a result, an interval can be quantified by two coordinated observers using forward projections alone, since $\Delta \bar{p} = \Delta {q}$ and $\Delta \bar{q} = \Delta {p}$.  Thus two coordinated observers can quantify intervals based only on their response to being influenced.

The length $\Delta p$ of a closed interval $[a,b]_{\P}$ along the chain $\P$ can be rewritten as
\begin{equation} \label{eq:length}
d([a,b]_{\P}) \equiv \Delta p = \frac{\Delta p + \Delta q}{2},
\end{equation}
since $\Delta p = \Delta q$ (Figure \ref{fig:coordination+distance+decomposition}A).

Next I introduce a valuation $D([[\P, \Q]])$ that quantifies the relationship between two coordinated chains $\P$ and $\Q$.  I require such a valuation to be based on projections of elements onto an observer chain.  This can be accomplished by selecting one element on each chain to define an interval, so that $D([[\P, \Q]]) \equiv D(\Delta p, \Delta q)$ where $\Delta p = p_b - p_a$, $\Delta q = q_b - q_a$ for  $a \in \P$ and $b \in \Q$.  Since there can be no preferred elements, I require the valuation to return the same quantity regardless of the elements selected to define the interval.  That is, I require that $D(\Delta p, \Delta q) = D(\Delta p', \Delta q')$ where $\Delta p' = p_{b'} - p_{a'}$, $\Delta q' = q_{b'} - q_{a'}$ for any ${a'} \in \P$ and any ${b'} \in \Q$.

I can consider multiple mutually-coordinated chains and show that they must be ordered \cite{Knuth+Bahreyni:EventPhysics}\cite{Bahreyni:Thesis}.  I can then consider intervals across chains and show that by joining them, they must obey associativity.  This leads again to the associativity equation, which forces our valuation $D$ to be additive under this process of joining, so that it is linear in $\Delta p$ and $\Delta q$.  This implies that $D(\Delta p, \Delta q) = k_p \Delta p + k_q \Delta q$ where $k_p$ and $k_q$ are constants to be determined.  One can show that since this must work for any pair of elements, we have that $k_p = -k_q$ leaving us with an arbitrary scale.  Choosing the scale to agree with that used for length (\ref{eq:length})
\begin{equation} \label{eq:distance}
D([[\P, \Q]]) = \frac{\Delta p - \Delta q}{2},
\end{equation}
where $\Delta p$ and $\Delta q$ are the projections of \emph{any} interval with one endpoint on each of two coordinated chains $\P$ and $\Q$.  This valuation, which quantifies the relationship between coordinated chains, is referred to as \emph{\textbf{distance}}.  This is illustrated in Figure \ref{fig:coordination+distance+decomposition}B.

Two coordinated observers define a two-dimensional subspace of the non-dimensional poset.  One dimension arises from the natural order of the poset; whereas the other dimension is induced by the fact that the coordinated chains can be ordered by projection \cite{Knuth+Bahreyni:EventPhysics}.  For this reason, the subspace is referred to as being 1+1 dimensional, where these two valuations, length and distance, each capture a different aspect of the 1+1-dimensional space.  Consider an interval (Figure \ref{fig:coordination+distance+decomposition}C) that forward projects to a closed interval on $\P$ with length $\Delta p$ and forward projects to a closed interval on $\Q$ with length $\Delta q$ resulting in the pair quantification $(\Delta p, \Delta q)$.  This pair can always be decomposed into a symmetric pair of lengths and an antisymmetric pair of distances
\begin{equation}
(\Delta p, \Delta q) = \Big(\frac{\Delta p + \Delta q}{2}, \frac{\Delta p + \Delta q}{2} \Big) + \Big(\frac{\Delta p - \Delta q}{2}, \frac{\Delta q - \Delta p}{2} \Big),
\end{equation}
which is referred to as the \emph{\textbf{symmetric-antisymmetric decomposition}} \cite{Knuth+Bahreyni:EventPhysics}.
Furthermore, the interval scalars corresponding to the pairs enjoy the same additive relation
\begin{equation} \label{eq:metric}
\Delta p \Delta q = \Big(\frac{\Delta p + \Delta q}{2} \Big)^2 - \Big(\frac{\Delta p - \Delta q}{2} \Big)^2,
\end{equation}
which provides a metric by relating the interval scalar to a length and a distance.  Last, in terms of projections onto coordinated chains, the pair transform (Figure \ref{fig:homogeneity}B) can be written as
\begin{eqnarray} \label{eq:pair-transform}
L_{\P \rightarrow \P'}(\Delta p, \Delta q)_{\P\Q} &=& (\Delta p', \Delta q')_{\P'\Q'} \\
 &=& \Big(\Delta p \, \sqrt{\frac{m}{n}}, \Delta q \, \sqrt{\frac{n}{m}}\Big)_{\P'\Q'} \nonumber
\end{eqnarray}
where closed intervals of length $\Delta p = \Delta q = k = \sqrt{mn}$ project to closed intervals of lengths $\Delta p' = m$ and $\Delta q' = n$.  This is analogous to the Bondi k-calculus formulation of special relativity \cite{Bondi:1980}.

These results constitute what I refer to as the \emph{\textbf{poset picture}}.  In the next subsection, I will summarize the relationship between the poset picture and spacetime.

%
%

\subsubsection{Emergent spacetime}
In this section I will show how the consistent quantification of a poset by an embedded observer gives rise to a concept of emergent spacetime, referred to as the \emph{\textbf{spacetime picture}}.

The symmetric-antisymmetric decomposition suggests a convenient change of variables:
\begin{equation} \label{eq:time+position}
\Delta t = \frac{\Delta p + \Delta q}{2} \qquad \mbox{and} \qquad \Delta x = \frac{\Delta p - \Delta q}{2}
\end{equation}
so that $\Delta p = \Delta t + \Delta x$ and $\Delta q = \Delta t - \Delta x$.
Any interval pair $(\Delta p, \Delta q)$ can be written as
\begin{equation}
(\Delta p, \Delta q) = (\Delta t, \Delta t) + (\Delta x, -\Delta x),
\end{equation}
where we refer to the two pairs on the right as the time and space components, respectively.  The interval scalar can be expressed similarly resulting in a quadratic form analogous to the Minkowski metric of flat spacetime
\begin{equation}
\Delta p \Delta q = {\Delta t}^2 - {\Delta x}^2,
\end{equation}
which results from the fact that the chains $\P$ and $\Q$ are coordinated, which is analogous to the concept of synchronized clocks.
Note that the signature of the metric is \emph{not arbitrary}, but is \emph{derived} and found to agree with the convention used in particle physics rather than the convention used in general relativity, where the time component carries the minus sign.  This is due to the antisymmetry of the antichain-like interval pair, which is the origin of the antisymmetry of space in differential geometry.  This is critical for another reason, since I will show in the next subsection that this is intimately related to the mass-energy relationship.

Similarly, the pair transformation (\ref{eq:pair-transform}) can be written as
\begin{equation}
\left(\Delta t' + \Delta x', \Delta t' - \Delta x' \right) =
\left((\Delta t + \Delta x) \sqrt{\frac{m}{n}}, (\Delta t - \Delta x) \sqrt{\frac{n}{m}} \right),
\end{equation}
which can be represented by a matrix multiplication. Solving for
$\Delta t'$ and $\Delta x'$ gives
\begin{eqnarray}
\Delta t' & = & \frac{\sqrt{\frac{m}{n}} + \sqrt{\frac{n}{m}}}{2} \Delta t + \frac{\sqrt{\frac{m}{n}} - \sqrt{\frac{n}{m}}}{2} \Delta x \\
\Delta x' & = & \frac{\sqrt{\frac{m}{n}} - \sqrt{\frac{n}{m}}}{2} \Delta t + \frac{\sqrt{\frac{m}{n}} + \sqrt{\frac{n}{m}}}{2} \Delta x.
\end{eqnarray}

By defining
\begin{equation} \label{eq:beta}
\beta = \frac{m-n}{m+n},
\end{equation}
and writing $1-\beta^2$, one can substitute these into the relations above and, with a little algebra, arrive at a Lorentz transformation analog
\begin{eqnarray}
\Delta t' & = & \frac{1}{\sqrt{1-{\beta}^2}} {\Delta t} + \frac{-{\beta}}{\sqrt{1-{\beta}^2}} {\Delta x} \\
\Delta x' & = & \frac{-{\beta}}{\sqrt{1-{\beta}^2}} \Delta t + \frac{1}{\sqrt{1-{\beta}^2}}  \Delta x.
\end{eqnarray}
These results suggest that time and space emerge as a uniquely consistent observer-based quantification of intervals in a poset.

The quantity $\beta$ in (\ref{eq:beta}), which can be written in terms of projections that amount to the ratio of a distance between chains over a length along chains
\begin{equation} \label{eq:speed}
\beta \equiv \frac{\Delta p - \Delta q}{\Delta p + \Delta q} = \frac{\Delta x}{\Delta t},
\end{equation}
emerges as the relevant quantity that relates two linearly-related chains that project to one another in a constant fashion.  In the special case of coordinated chains where $m=n$, we have that $\beta = 0$ so that in the spacetime picture coordinated observers are at rest with respect to one another.  Extreme values of $\beta = \pm1$ correspond to the cases where $n=0$ and $m=0$, respectively, and thus to projection-like intervals.  Since the interval scalar is invariant among linearly-related chains (constant projection), extremal values of $\beta$ with respect to one observer chain are extremal with respect to all other linearly-related chains.  Thus in the spacetime picture, there is a maximum invariant speed.\footnote{The fact that there is a maximum invariant speed is a consequence of the derivation---not an assumption.} However, in the poset picture, there is no space and no motion---only relationships.

\begin{figure}[t]
  \begin{center}
  \includegraphics[height=0.15\textheight]{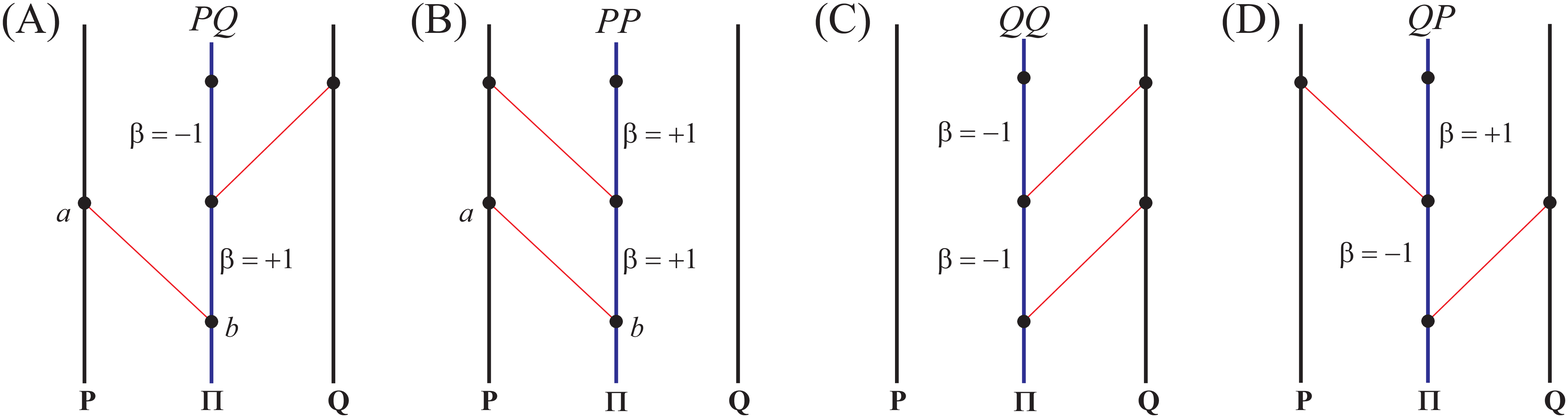}
  \end{center}
  \caption{(A)-(D) Illustrates a free particle $\Pi$ influencing one of two coordinated observers $\P$ or $\Q$ twice with influence sequences represented by the symbol strings $PQ$, $PP$, $QQ$, $QP$, respectively.  We first note that since each influencing event affects only one observer chain, it is not possible to consistently assign a position and a time for that event without projecting through one of the subsequent events that projects to the other chain.
  }
  \label{fig:helicity}
\end{figure}

\subsection{Observer-Based Description of the Free Particle}
Traditionally, a free particle is a particle that is free from all external forces.  This, of course, is an unattainable idealization.  However, such idealizations are important in developing a theory.  Here, I similarly define a \emph{\textbf{free particle}} to be a particle that influences others, but is not itself influenced.  The possibility that a free particle can influence others is critical; since if it did not, it would not detectable by any observer and could not be described or quantified.  A free particle is modeled as a chain of events, each of which represents an act of influence directed toward some other particle.  I will formulate a relevant description of a free particle based on its influence on two coordinated observers, which implicitly define a 1+1 dimensional emergent spacetime.\footnote{There are some technical difficulties here that should be acknowledged since it appears that they lie at the heart of the discord between spacetime physics and quantum mechanics as described here.  The two observers cannot be precisely coordinated when obeying the influence-based connectivity rules since they cannot simultaneously influence one another while being influenced by the free particle.  One can coordinate them in a coarse-grained sense by having them influence one another both before and after the free particle is studied, and ensuring that they agree on the lengths of these larger intervals.  As I will discuss below there is a fundamental limit to spatial and temporal resolution, and both coordination and the ability to determine whether a particle is a member of the 1+1 dimensional subspace is subject to this limit.}

Figures \ref{fig:helicity}(A-D) illustrate a free particle $\PI$ performing two successive acts of influence each affecting one of two coordinated observer chains $\P$ or $\Q$.  I refer to an act of influence directed toward $\P$ as a \emph{\textbf{P-move}}, and an act of influence directed toward $\Q$ as a \emph{\textbf{Q-move}}.  This allows me to express these influence patterns as sequences with the symbol strings PQ, PP, QQ, QP, for Figures \ref{fig:helicity}A-D, respectively.  These influence patterns can be mapped to bit strings that encode the particle's behavior.  More specifically, the observers $\P$ and $\Q$ can only record whether or not the particle influenced them at any event $x$ along their respective chains.  This represents the only information made available to the observers to describe the free particle.  In the following subsections, I will consider an observer-based description of the free particle's behavior in terms of events, intervals and rates.

\subsubsection{Influence Sequences and Spacetime Paths}
\begin{figure}[t]
  \begin{center}
  \includegraphics[height=0.30\textheight]{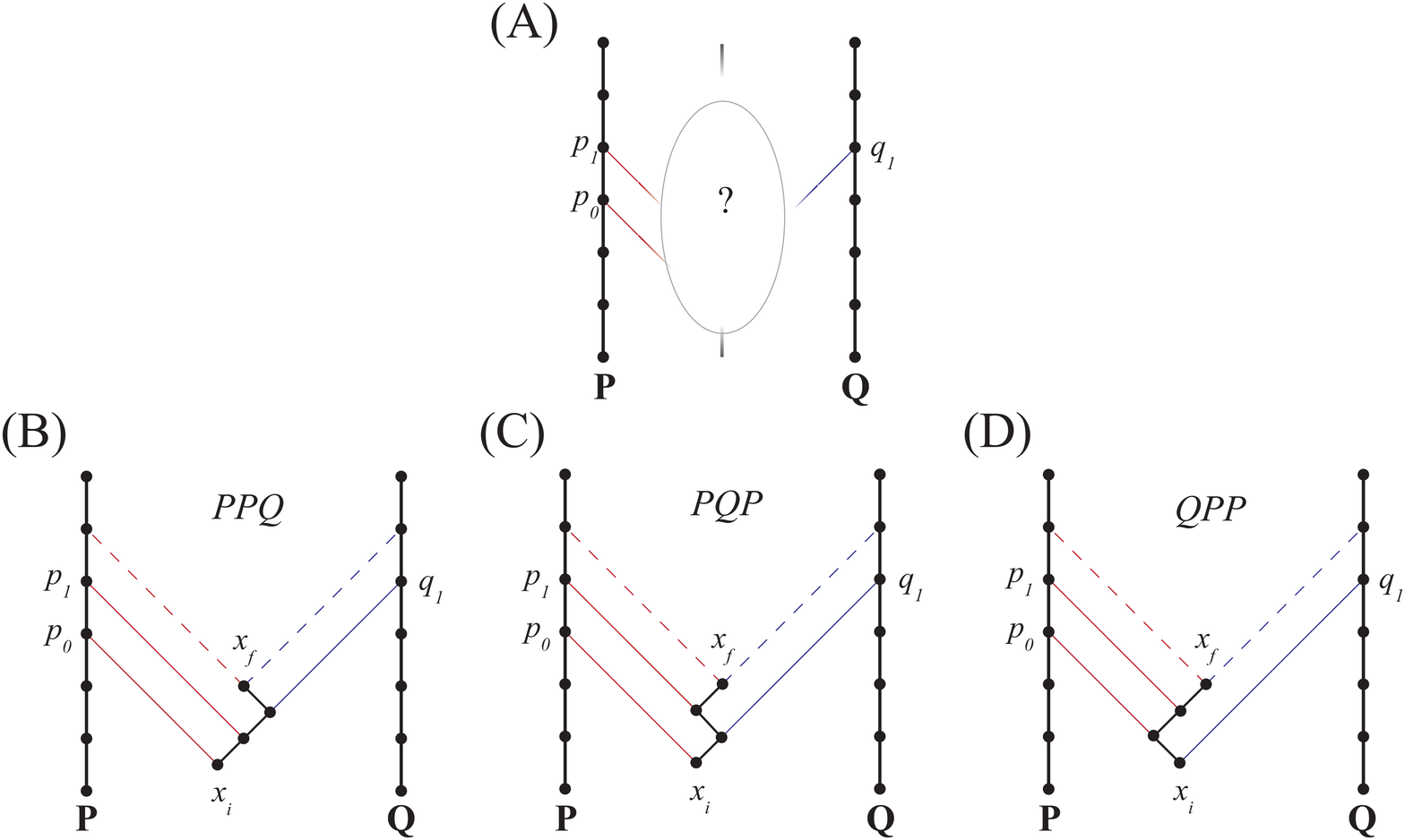}
  \end{center}
  \caption{(A) The observer chains $\P$ and $\Q$ record influence events $p_0, p_1$ and $q_1$ from the free particle.  However, the relative ordering of the P events and Q events cannot be determined. (B)-(D) Illustration of the three possible influence sequences.  The particle chain is depicted in a poset diagram that is readily translated into the spacetime picture, where the particle zig-zags back and forth at the maximum speed.  For $P = 2$ and $Q = 1$ there are three possible sequences, $PPQ, PQP, QPP$, each of which corresponds to a spacetime path.  The dashed lines at the top of the particle chain indicate potential successive influence events.}
  \label{fig:quantum-projections---zigzags}
\end{figure}

Imagine that observer chain $\P$ records influence events at $p_1, p_3$ and $p_4$; whereas observer chain $\Q$ records influence events at $q_2$ and $q_5$.  The fact that the events recorded by $\P$ are elements of a different ordered chain than the events recorded by $\Q$ implies that they are incomparable and thus cannot be mutually ordered.  This is a subtle, but important point.  The events representing acts of influence along the particle chain $\PI$ are ordered.  However, the events representing response to such influence experienced by different particle chains \emph{cannot} be ordered.  It is not a matter of missing information---there is no way to relate $p_1$ and $q_2$.

This poses difficulties when assigning spacetime coordinates to events along the particle chain based on events representing responses to influence along the observer chains. I will show later, when discussing inference, that all possible sequences must be considered. More specifically, given that there were $P = 3$ P-moves and $Q = 2$ Q-moves, this means that there are
\begin{equation} \label{eq:number-of-sequences}
N = \binom{P + Q}{P} = \binom{P + Q}{Q} = \frac{(P + Q)!}{P! Q!}
\end{equation}
or $N = 10$ possible influence sequences of the form $PPPQQ, PPQPQ, PPQQP$, \ldots, $QQPPP$.

The indeterminacy of the influence sequence in the poset picture translates to an indeterminacy of the spacetime path assigned to the particle by the observers in the spacetime picture.  It is not that it is impossible to \emph{determine} the path that the particle took, but rather there is \emph{no definable path} based on the events recorded by the observers.  Later, I will show that all possible sequences, or paths, must be considered when making inferences about the particle's behavior.

These concepts are illustrated in the case of $P = 2$ P-moves and $Q = 1$ Q-moves in Figure \ref{fig:quantum-projections---zigzags}.

\subsubsection{Events: Positions and Times} \label{sec:events-positions-times}
Recall that time intervals and spatial distances are found by projecting intervals onto both the $\P$ and $\Q$ observer chains and computing $\Delta t$ and $\Delta x$ by (\ref{eq:time+position}).
The discrete nature of the events results in a set of discrete spacetime coordinates.  Since at minimum one will have either $\Delta p = 1, \Delta q = 0$ or $\Delta p = 0, \Delta q = 1$, the time coordinate will advance with a minimum time step of $\Delta t = +1/2$ units and the space coordinate will either advance to the left or right with $\Delta x = \pm 1/2$.  This results in a chessboard-like grid where the particle is constrained to make bishop-moves without being able to change square color.  This fundamental limitation on measuring the spacetime position of a particle is reminiscent of the Compton wavelength given by $\lambda = \frac{h}{Mc}$, which is dependent on mass.

The effect of this limitation can be seen by considering the projection-like interval $[a,b]$ indicated by the labeled events in both Figures \ref{fig:helicity}A and B.  Clearly $\Delta p = 0$ for this interval.  However, to obtain a value for $\Delta q$ one must project the interval onto chain $\Q$.  Not only must the event $a$ project from one observer chain to the other, but also the projection of the element $b$ depends on the subsequent influence of $\PI$ on the chain $\Q$ as indicated by the fact that in Figure \ref{fig:helicity}A the event $b$ forward projects to $\Q$ via the next act of influence; whereas in Figure \ref{fig:helicity}B, the event $b$ has yet to project to $\Q$.  The result is that the position and time assigned to any event along the particle chain depends on successive acts of influence.

\begin{figure}[t]
  \begin{center}
  \includegraphics[height=0.20\textheight]{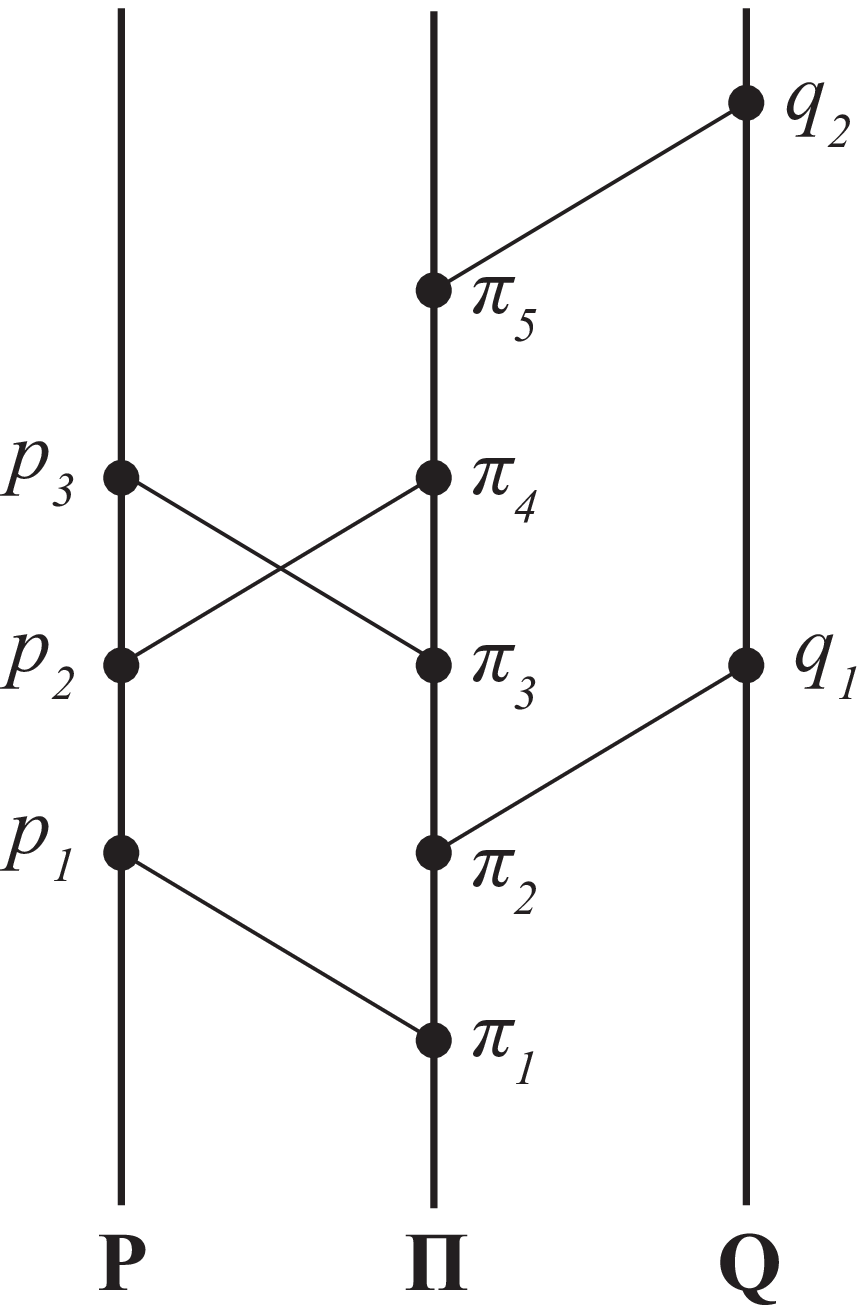}
  \end{center}
  \caption{This illustrates how by influencing the particle, one can, in principle, partially order some of the recorded influence events.  In this example, the event $p_2$ on $\P$ influences the particle $\PI$ at event $\pi_4$, which precedes $\pi_5$, which in turn influences $q_2$ on $\Q$. As a result, the events $p_2$ and $q_2$ are ordered such that $p_2 < q_2$.  This could potentially serve as a model for a measurement process, where some events along $\P$ and $\Q$ can be ordered at the expense of influencing the particle.
  }
  \label{fig:measurement}
\end{figure}

Despite the difficulties in ordering events representing observer responses to influence and thus assigning spacetime coordinates, it is possible that by influencing the particle itself, one could define such an ordering across observer chains.  One such possibility is illustrated in Figure \ref{fig:measurement} where event $p_2$ represents the observer $\P$ influencing the particle chain $\PI$ at event $\pi_4$, which in turn precedes $\pi_5$ where the chain $\PI$ influences observer $\Q$ at $q_2$.  The result is that $p_2$ and $q_2$ are ordered such that $p_2 < q_2$.  This could potentially serve as a model for a measurement process, where some events along $\P$ and $\Q$ can be ordered at the expense of influencing the particle.

\subsubsection{Intervals: Helicity and Zitterbewegung}
We now consider quantifying intervals between successive events along the free particle chain.  As described above, the interval consisting of an event on the chain $\PI$ and its successor form a projection-like interval with respect to the observer chains.  That is, each interval defined by pairs of successive events along the free particle chain has either a projection of $\Delta p = 0$ or $\Delta q = 0$.  From the expression for the parameter $\beta$ (\ref{eq:speed})
\begin{equation}
\beta \equiv \frac{\Delta p - \Delta q}{\Delta p + \Delta q} = \frac{\Delta x}{\Delta t},
\end{equation}
we find that $\beta = \pm 1$.  That is, the observers interpret the free particle as zig-zagging back-and-forth at the maximum invariant speed, which is analogous to the speed of light.  This is a quantum effect, known as \emph{\textbf{Zitterbewegung}} \cite{Huang:1952zitterbewegung}\cite{Hestenes:1990zitterbewegung}\cite{Hestenes:2008electron}.  It was first proposed by Schr\"{o}dinger in 1930, and only recently observed in the laboratory \cite{Catillon+etal:2008clock}\cite{Gerritsma+etal:2010quantum}.  The phenomenon arises from the fact that the speed eigenvalues of the Dirac equation are $\pm c$ (the speed of light) \cite{Schrodinger1930:kraftefreie}.

The sign of $\beta$ assigned to an interval consisting of a pair of successive influence events is dictated by the first act of influence.  That is, if the first act of influence is a P-move, then $\beta = +1$ so that the particle is interpreted as moving away from the chain $\P$.  Otherwise, for a Q-move, $\beta = -1$ so that the particle moves away from $\Q$.  This directionality is referred to as \emph{\textbf{helicity}}, which is thought to be the 1+1 dimensional analogue of spin.

Another consequence is that there are two ways to arrive at a particular spacetime coordinate, each depending on the direction of the previous act of influence.  As a result, an initial spacetime state actually consists of two possible helicity states that must be accounted for when making inferences about the particle's behavior.  Furthermore, since there are only two ways to arrive at a particular spacetime coordinate, we see that in this model, there can be a maximum of two particles occupying a particular spacetime position each with different helicity states, which is analogous to the \emph{\textbf{Pauli exclusion principle}}.

\subsubsection{Rates: Mass, Energy and Momentum}
Instead of focusing the description of a particle on intervals, one can instead consider rates.  The two are Fourier transforms of one another, and the corresponding relevant variables will share this relationship.  Rates differ from intervals in that they are well-defined only over a range of events; so that given a single event, one cannot determine a rate.  Consider an influence sequence of a large number of events $N$.  Since the P's and Q's cannot be mutually ordered, specific patterns and correlations in the sequence cannot be considered.  This limits quantification to simply counting the number of P and Q events.  Given a closed interval consisting of $N$ events along the particle chain $\PI$, one can project this interval onto the observer chains resulting in the pair quantification $(\Delta p,\Delta q)$ and define average rates $r_P$ and $r_Q$ as
\begin{equation}
r_P = \frac{N}{\Delta p} \qquad \mbox{and} \qquad r_Q = \frac{N}{\Delta q}.
\end{equation}

Given that ${\Delta p}{\Delta q}$ is invariant, the product of these rates is also invariant
\begin{equation} \label{eq:rate-equation}
{r_P}{r_Q} = \frac{N^2}{{\Delta p}{\Delta q}}.
\end{equation}
By applying the symmetric-antisymmetric decomposition, one obtains two rate-dependent parameters that are conjugate to time and space
\begin{equation} \label{eq:rate-sym-asym}
{r_P}{r_Q} = \Big({\frac{r_P + r_Q}{2}}\Big)^2 - \Big({\frac{r_Q - r_P}{2}}\Big)^2
\end{equation}
where I have harmlessly reversed the sign inside the square of the second term for reasons that will be explained.
These quantities are analogous to mass, energy, and momentum where the geometric mean of the rates $r_P$ and $r_Q$
\begin{equation}
M := \sqrt{{r_P}{r_Q}}
\end{equation}
is analogous to the \emph{\textbf{mass}},
the arithmetic mean of the rates
\begin{equation}
E := \frac{r_P + r_Q}{2}
\end{equation}
is analogous to the \emph{\textbf{energy}},
and the half-difference
\begin{equation}
p := \frac{r_Q - r_P}{2}
\end{equation}
is analogous to the \emph{\textbf{momentum}}
so that (\ref{eq:rate-sym-asym}) is analogous to the energy-momentum relation
\begin{equation} \label{eq:mass-energy-momentum}
M^2 = E^2 - p^2.
\end{equation}
Keep in mind that only two parameters are necessary to describe the rates at which the particle influences others.  In the poset picture, these are simply $r_P$ and $r_Q$; whereas in the spacetime picture one requires any two of $M$, $E$, or $p$, which are related by (\ref{eq:mass-energy-momentum}).

I will now show that these parameters in the spacetime picture obey additional expected relations.  First, I verify that the speed, $\beta$, is given by the ratio of the momentum to the energy
\begin{equation}
\beta = \frac{\Delta x}{\Delta t} = \frac{\Delta p - \Delta q}{\Delta p + \Delta q} = \frac{\frac{N}{\Delta q} - \frac{N}{\Delta p}}{\frac{N}{\Delta p} + \frac{N}{\Delta q}} =  \frac{p}{E}.
\end{equation}
The parameters $E$ and $p$, rather than being properties of the particle, are descriptions of its behavior, and as such, their values change as one transforms from one observer chain to another.  In the case of linearly-related observer chains (\ref{eq:pair-transform}), rates transform inversely to intervals
\begin{equation}
{r_P}' = \sqrt{\frac{n}{m}} r_P \qquad \mbox{and} \qquad {r_Q}' = \sqrt{\frac{m}{n}} r_Q
\end{equation}
so that their product is invariant as expected.
The transformed quantities $E'$ and $p'$ can be written in terms of the rates quantified by the original chain as
\begin{equation} \label{eq:E'-P'}
E' = \frac{{r_P}' + {r_Q}'}{2} = \frac{1}{2}\Big(\sqrt{\frac{n}{m}} r_P + \sqrt{\frac{m}{n}} r_Q \Big) \quad \mbox{and} \quad p' = \frac{{r_Q}' - {r_P}'}{2} = \frac{1}{2}\Big(\sqrt{\frac{m}{n}} r_Q - \sqrt{\frac{n}{m}} r_P \Big).
\end{equation}
In both cases, by both adding and subtracting $\sqrt{\frac{m}{n}}\frac{r_P}{4}$ and $\sqrt{\frac{n}{m}}\frac{r_Q}{4}$ and regrouping terms, one can rewrite $E'$ and $p'$ in (\ref{eq:E'-P'}) as
\begin{equation}
E' = \gamma E + \beta \gamma p  \qquad \mbox{and} \qquad p' = \beta \gamma E + \gamma p
\end{equation}
where $\beta = \frac{m-n}{m+n}$ as in (\ref{eq:pair-transform-one-chain}) and (\ref{eq:pair-transform}), and $\gamma = (1-\beta^2)^{-1/2} = \frac{1}{2}(\sqrt{\frac{m}{n}} + \sqrt{\frac{n}{m}})$.  When the particle has momentum $p = 0$ with respect to the first chain, so that $r_P = r_Q$, then $E=M$, which results in the familiar relations
\begin{equation}
E' = \gamma M  \qquad \mbox{and} \qquad p' = \beta \gamma M.
\end{equation}


Note that the sign of the momentum is such that a particle that influences more often to the right than to the left in the poset picture corresponds to a particle moving to the left with a negative $\beta$ in the spacetime picture.  A particle at rest in the spacetime picture influences chains $\P$ and $\Q$ with equal probabilities.

Now I have not shown that these quantities are conserved in a multi-particle setting, which is crucial for a complete development.  In addition, since influence events define rates, which are related to momentum and energy, one would expect that influence is related to force.  These topics will be explored in future work.  At present, I am  simply focused on obtaining a complete and accurate single particle model.  What is satisfying is that this model makes an explicit connection between rates on the one hand and mass, energy and momentum on the other, in a way that is not unlike the internal electron clock rate first hypothesized by de Broglie in his 1924 thesis \cite{Hestenes:2008electron}.

Furthermore, the rate at which a particle influences others dictates the smallest resolution with which it can be localized.  It is in this sense that this resolution is expected to be related to mass, which further supports the apparent analogy between this resolution and the Compton wavelength.  Another way to conceive of this is that the overall rate of particle influence, or mass, is responsible for generating its spacetime positions.

\section{Inference} \label{sec:Inference}
In this section, I consider optimal inferences that observers can make about a free particle.  The process of inference involves ranking logical statements that can be made about a system based on all available information.  I will summarize how quantification of statements under the constraint of associativity leads to the sum and product rules of probability theory \cite{Knuth&Skilling:2012}.

In typical applications, the states of the system being discussed are in some sense mutually exclusive.  However, here inferences will be made about measurement sequences.  Measurement sequences can be related to one another via algebraic operations obeying associativity and distributivity.  I will summarize recent work that has shown that consistent assignment of probabilities in this case requires the quantification of measurement sequences.  Quantifying sequences with pairs of real numbers results in a set of constraint equations that can be identified as the Feynman rules for manipulating quantum amplitudes \cite{GKS:PRA}\cite{GK:Symmetry}.

I show that making consistent inferences about a free particle that influences two coordinated observers results in the Feynman checkerboard model, which is known to lead to the Dirac equation in 1+1 dimensions.  This will be accomplished by \emph{deriving} the amplitude assignments \cite{Knuth:fermions} suggested by Feynman and Hibbs \cite{Feynman&Hibbs}.  The result is that the Dirac equation is viewed as a recipe for making optimal inferences about the behavior of a free particle that influences others.

\subsection{Quantifying Statements: Probability Theory}
In this section, I consider the set of statements that one can make about a system described by a set of states, and summarize previous work on the consistent quantification of such statements \cite{Knuth:laws}\cite{Knuth:duality}\cite{Knuth:measuring}\cite{Knuth&Skilling:2012}, which results in probability theory.

\subsubsection{States and Statements}
Consider a system that can be described as being in one of a finite number $N$ of states labeled $a_1, a_2, \ldots, a_N$.  In addition, consider an observer who possesses a possibly imperfect state of knowledge about the state of the system.  Very crudely, one can describe such a state of knowledge as being represented by a set of potential states that the system is thought to be in.  That is, an observer possessing no information about the state of the system could represent his or her state of knowledge in terms of the set of all $N$ possible states.  On the other extreme, an observer knowing the state of the system precisely would represent his or her state of knowledge in terms of a set consisting of the actual state of the system.  This crude description of a state of knowledge in terms of potential states allows one to describe the space of all possible states of knowledge as the \emph{\textbf{power set}} of states, which is the set of all subsets of the $N$ states.  Each element of this space is a set of potential states, referred to as a \emph{\textbf{statement}}.

Statements in this power set are naturally ordered, generically denoted $\leq$, in terms of subset inclusion $\subseteq$, which is isomorphic to logical implication $\rightarrow$, thus forming a \emph{\textbf{Boolean lattice}}.  For example, the statement that the system is in one of the states given by the set $\{a_1, a_2\}$ implies the statement that the system is in one of the states $\{a_1, a_2, a_3\}$.  Not all statements are comparable in this way, since not all sets of potential statements are subsets of one another.  Furthermore, singleton sets, which express a state of complete knowledge, are mutually exclusive and form the atoms from which all other nonempty sets can be constructed by set union.

For every pair of statements, there exists a least upper bound called the join, denoted $\JOIN$, found by taking the set union $\cup$ of the sets, which is isomorphic to the logical $\OR$ operation.  Dually, there also exists a greatest lower bound called the meet, denoted $\MEET$, found by taking the set intersection $\cap$, which is isomorphic to the logical $\AND$ operation.    One equivalently can view this as an order-theoretic structure resulting in a hierarchy of statements ordered by implication, or as an algebra of operations defined on a set of statements that obey certain symmetries.

\subsubsection{Quantification of a Boolean Lattice}
This section focuses on the consistent quantification of the relationship between ordered pairs of statements in the Boolean lattice so that statements can be compared and ranked.  Such ordered pairs of statements can be viewed as intervals analogous to intervals of events in Section \ref{sec:spacetime}.  However, unlike a general poset, the Boolean lattice possesses a great of deal of symmetry, which is sufficient to constrain quantification.  Rather than deriving the results in detail, I will outline how symmetries result in functional equations and direct the interested reader to the following papers \cite{Knuth:laws}\cite{Knuth:duality}\cite{Knuth:measuring}\cite{Knuth&Skilling:2012}, and especially \cite{GK:Symmetry}, which focuses on the relationship between probability theory and quantum mechanics.

Since statements are ordered by implication, a valuation consistent with implication encodes \emph{\textbf{degrees of implication}}.  The bi-valuation $w$, is a functional that takes an ordered pair of statements $y$ and $z$, or equivalently the interval $[y,z]$, to a real number $w(y,z)$, such that if $x \leq y$ then $w(x,z) \leq w(y,z)$.  The valuation $w(y,z)$ can be read as `\emph{the degree to which $z$ implies $y$}', where the second argument $z$ is referred to as the \emph{\textbf{context}}.

The first symmetry employed is associativity of the $\JOIN$ operation (logical $\OR$) with respect to the first argument of the bi-valuation.  As described in detail in Section \ref{sec:chain-quantification}, without loss of generality, this results in additivity.  For mutually exclusive statements $x$ and $y$, where $x \MEET y = \Emptyset$, which is the null set, the bi-valuations are additive so that
\begin{equation} \label{eq:bivaluation-sum}
w(x \JOIN y, z) = w(x,z) + w(y,z).
\end{equation}
Next, consider chaining context, which is performed by joining intervals that share a single endpoint.  Again, similar to the derivation in Section \ref{sec:chain-quantification}, intervals can be joined in an associative fashion, which results in additivity.  However, given that I selected additivity for the $\JOIN$ operation in (\ref{eq:bivaluation-sum}), it can be shown that by considering distributivity of $\MEET$ over $\JOIN$ the only invertible transform of additivity consistent with this is the multiplicative product so that for $x \leq y \leq z$,
\begin{equation} \label{eq:bivaluation-chaining}
w(x,z) = w(x,y) \, w(y,z),
\end{equation}
which is referred to as the chain rule \cite{Knuth&Skilling:2012}.\footnote{Another approach described in \cite{GK:Symmetry} shows that associativity of the direct product of different statement spaces and distributivity of the direct product over $\JOIN$ results in multiplication.  The chain rule (\ref{eq:bivaluation-chaining}) then follows as a special case.}
By considering more general cases, it can be shown that these constraint equations that ensure consistent bi-valuation assignments are the familiar sum and product rules of probability theory \cite{Knuth:duality}\cite{Knuth:measuring}\cite{Knuth&Skilling:2012}\cite{GK:Symmetry}.  Changing notation by denoting the bi-valuation with $Pr$ and employing a solidus instead of a comma to separate the arguments results in\footnote{I have omitted the product rule for the direct product, which combines hypothesis spaces, mentioned in the previous footnote.}
\begin{equation} \label{eq:probability-sum}
Pr(x \JOIN y | z) = Pr(x|z) + Pr(y|z) - Pr(x \MEET y|z)
\end{equation}
\begin{equation} \label{eq:probability-prod}
Pr(x \MEET y| z) = Pr(x|y \MEET z) \, Pr(y|z).
\end{equation}
In arriving at (\ref{eq:bivaluation-chaining}) there was a freedom of overall scale, which was set so that the bi-valuations have a maximum value of unity.  Thus in the case where a statement $x$ implies a statement $z$ ($x \rightarrow z$, or more generally $x \leq z$) we have that $Pr(z|x) = 1$, whereas $0 < Pr(x|z) \leq 1$.  At the other extreme, the relationship between two mutually exclusive statements, such as $x$ and $y$ where $x \MEET y = \Emptyset$, is quantified by $Pr(x|y) = Pr(y|x) = 0$.

The result is a scalar quantification of the relationship among logical statements ordered by implication.  In this sense, probability is a real-valued quantification of logical implication.

\subsection{Inferences about a Free Particle}
I now consider inferences that two coordinated observers could make about the behavior of a free particle.
This consists of the task of computing the probability of recording $P$ P-moves and $Q$ Q-moves.  One might be tempted to simply sum the probabilities, or marginalize, over each of the possible, or hypothetical, ordered sequences.  However, there are no ordered sequences, and thus it is meaningless to make statements about them, much less to quantify such statements with probabilities.

Instead, one must assign a probability to the unordered set of recorded influence events.  This lies at the heart of the issue as to when one can use probability theory directly, and when one must resort to quantum mechanical amplitudes to compute probabilities before probability theory is applied.  Feynman and Hibbs \cite{Feynman&Hibbs} distinguish between \emph{exclusive alternatives} where the possible situations can be unambiguously distinguished, and \emph{interfering alternatives} where it is impossible to distinguish between the possible situations.  They note that probabilities can be directly assigned in the case of exclusive alternatives, but quantum amplitudes must be used to compute probabilities in the case of interfering alternatives.  
Here one can see precisely in what sense the alternatives cannot be distinguished.  Given an unordered set of influence events, ordered sequences of influence events do not exist; thus nothing can be said about them.  Furthermore, the assumption that something can be said about them, as is made in classical mechanics, assumes too much and results in wrong answers \cite{GK:Symmetry}.

\subsection{Quantifying Sequences: Quantum Mechanics}
Unordered sets of influence sequences are clearly related to the set of all possible ordered sequences.  The assignment of probabilities in both cases must be consistent with this relationship.  I review our previous work, which considers the quantification of measurement sequences in the spacetime picture \cite{GKS:PRA, GKS:me09, GK:Symmetry} and briefly describe how this leads to the Feynman rules for manipulating quantum amplitudes. I will then proceed to consider quantification of influence sequences in the poset picture by insisting on consistency with the quantification of measurement sequences in the spacetime picture.

\subsubsection{Quantifying Measurement Sequences}
Consider a quantum mechanical system that is subject to a sequence of measurements $M_1, M_2, M_3, \ldots$, where each measurement $M_i$ has potential outcomes denoted by $m_i, {m_i}', {m_i}'', \ldots$.  A given experiment yields a sequence of outcomes, such as $\lseq{m_1}{{m_2}''}{{m_3}'}$.  One can also consider the possibility that measurements can be coarse-grained, such that a subset of measurement outcomes are not distinguished.  For example, consider that measurement $M_2$ is performed in such a way that outcomes ${m_2}'$ and ${m_2}''$ are not distinguished from one another.  In such a case, the outcome of the experiment is expressed as the sequence $\lseq{m_1}{({m_2}', {m_2}'')}{m_3}$, where $({m_2}', {m_2}'')$ indicates a coarse-grained measurement.

There are two possible ways to order sequences of measurement outcomes based on the identification of subsequences.  First, given a sequence of $N$ measurement outcomes, one can select some contiguous subsequence of length $M<N$, such that the longer sequence contains the shorter sequence.  For example, the sequence $\lseq{m_1}{{m_2}''}{{m_3}'}$ contains the subsequence $\seq{m_1}{{m_2}''}$.  This allows one to define an algebraic operation that combines two measurement sequences in \emph{\textbf{series}} if the final measurement of one sequence is the same as the initial measurement of the other.  That is, if $A = \seq{m_1}{m_2}$ and $B = \seq{m_2}{m_3}$, then one can write
\begin{equation} \label{eq:series}
C = A \mydot B = \seq{m_1}{m_2} \mydot \seq{m_2}{m_3} = \lseq{m_1}{m_2}{m_3}.
\end{equation}
The concept of coarse-graining provides another way of ordering sequences by selecting a subsequence.  One can say that the sequence $\lseq{m_1}{({m_2}', {m_2}'')}{m_3}$ contains both the sequence $\lseq{m_1}{{m_2}'}{m_3}$ and the sequence $\lseq{m_1}{{m_2}''}{m_3}$.  This allows one to define an algebraic operation that combines sequences in \emph{\textbf{parallel}} so that if $A = \lseq{m_1}{{m_2}'}{m_3}$ and $B = \lseq{m_1}{{m_2}''}{m_3}$ and ${m_2}' \neq {m_2}''$ then
\begin{equation}
C = A \vee B = \lseq{m_1}{{m_2}'}{m_3} \vee \lseq{m_1}{{m_2}''}{m_3} = \lseq{m_1}{({m_2}', {m_2}'')}{m_3}.
\end{equation}
These algebraic operations of series, denoted $\mydot$, and parallel, denoted $\vee$, obey certain symmetries that constrain quantification of the sequences.  For example, $\mydot$ is associative; whereas $\vee$ is both commutative and associative.  Furthermore, $\mydot$ is both left- and right-distributive over $\vee$.

In \cite{GKS:PRA, GKS:me09, GK:Symmetry}, we consider quantification of sequences with pairs of real numbers called amplitudes.  Consider sequences $A$ and $B$ quantified by real-valued pairs $\mathbf{a} = (a_1, a_2)$ and $\mathbf{b} = (b_1, b_2)$, respectively. Under the parallel operation, there must be some function $\oplus$ that relates the amplitudes $\mathbf{a}$ and $\mathbf{b}$ to the amplitude $\mathbf{c}$ quantifying the sequence $C = A \vee B$ so that
$\mathbf{c} = \mathbf{a} \oplus \mathbf{b}$
where $\oplus$ obeys the symmetries of $\vee$: commutativity and associativity.  Similarly, under the series operation, there must be some function $\odot$ that relates the amplitudes $\mathbf{a}$ and $\mathbf{b}$ to the amplitude $\mathbf{c}$ quantifying the sequence $C = A \mydot B$ so that
$\mathbf{c} = \mathbf{a} \odot \mathbf{b}$
where $\odot$ obeys the symmetries of $\mydot$: associativity, and left- and right-distributivity over $\oplus$.

Since $\oplus$ is both commutative and associative, it is constrained to be pair-wise additive so that
\begin{equation}
\mathbf{a} \oplus \mathbf{b} \equiv \begin{pmatrix}a_1\\a_2\end{pmatrix} \oplus \begin{pmatrix}b_1\\b_2\end{pmatrix} = \begin{pmatrix}a_1+b_1\\a_2+a_2\end{pmatrix}.
\end{equation}
Distributivity of $\odot$ over $\oplus$ constrains the combination of pairs to be a bilinear multiplicative form.  Associativity of $\odot$ further constrains the solution to five possible multiplicative forms (three commutative and two non-commutative forms).  A further requirement that the probability of a sequence be a function of the measurement outcomes such that $P(\lseq{m_1}{m_2}{m_3}) = Pr(m_2, m_3 | m_1)$, and that probability obeys the product rule when combining sequences in series so that $Pr(m_2, m_3 | m_1) = Pr(m_3 | m_2) Pr(m_2 | m_1)$ puts further constraints that eliminate the two non-commutative forms of multiplication for $\odot$.  Different arguments to obtain a unique form for $\odot$ are made in \cite{GKS:PRA} and \cite{GK:Symmetry}, which is found to be complex multiplication
\begin{equation}
\mathbf{a} \odot \mathbf{b} \equiv \begin{pmatrix}a_1\\a_2\end{pmatrix} \odot \begin{pmatrix}b_1\\b_2\end{pmatrix} = \begin{pmatrix}a_1b_1 - a_2b_2\\a_1b_2+a_2b_1\end{pmatrix}
\end{equation}
where the probability is given by the Born rule where for sequence $A = \seq{m_1}{m_2}$ we have that
\begin{equation}
P(A) = Pr(m_2|m_1) = {a_1}^2 + {a_2}^2.
\end{equation}
As a result, the Feynman rules for computing with quantum amplitudes is the uniquely consistent way of computing probabilities of measurement sequences \cite{GKS:PRA, GKS:me09, GK:Symmetry}.

\subsubsection{Quantifying Influence Sequences}
The quantification of measurement sequences described above was originally formulated with the spacetime picture in mind.  Here I consider the quantification of influence sequences defined by influence events along the $\P$ and $\Q$ observer chains, and require that this be consistent with the quantification of the corresponding measurement sequences in the spacetime picture.

\begin{figure}[t]
\begin{center}
  \includegraphics[height=0.18\textheight]{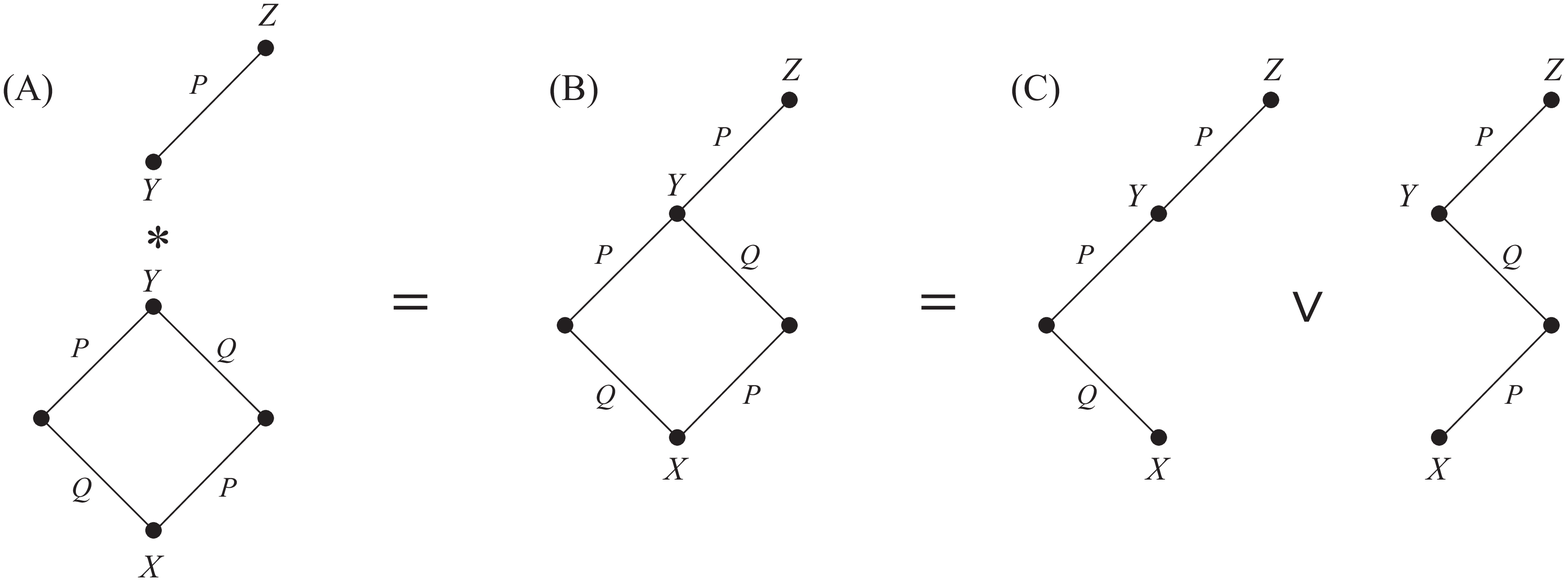}
\end{center}
  \caption{This figure illustrates the decomposition of the sequence $[(P,Q),P]$ into (A) $[(P,Q)] * [P]$ and its relation to the decomposition of the sequence (B) $[(P,Q),P]$ into (C) $[Q,P,P] \vee [P,Q,P]$. To maintain consistency, all three cases must by assigned the same amplitude.  The result is that unordered initial states $[(P,Q)]$ in (A) are handled by propagating a pair of amplitudes, which is analogous to a Pauli spinor.}
  \label{fig:diamond-decomposition}
\end{figure}

Consider the situation where a particle is prepared in an initial spacetime state $Y$.  The particle then performs a $\move{P}$-move arriving at a final state $Z$ (Figure \ref{fig:diamond-decomposition}A top).  In the spacetime picture, the amplitude is assigned to the measurement sequence $[Y,Z]$.  However, in the poset picture, this process is associated with an influence sequence consisting of a single influence event $[P]$ where the $\move{P}$-move serves to propagate the particle from $Y$ to $Z$.

Na\"{i}vely, one may be inclined to impose consistency by assigning the amplitude associated with the measurement sequence $[Y,Z]$ to the influence sequence $[P]$. However, in doing so, one can show that such quantification fails to capture relevant details about the influence sequences.  This is because it neglects the fact that influence sequences are measurement sequences in their own right. To make inferences about them, one must assign probabilities to pairs of influence events, which is performed by assigning amplitudes to $[P,P], [P,Q], [Q,P]$, and $[Q,Q]$.  Performing inference in the spacetime picture by assigning amplitudes to measurement sequences $[Y,Z]$ neglects the fact that the initial state $Y$ may be ambiguous since the particle may have arrived at $Y$ by either a $\move{P}$- or $\move{Q}$-move.  Another way of saying this is that \emph{\textbf{helicity matters}}!

The procedure with which one assigns an amplitude to an influence sequence such as $[P]$ should be able to accommodate both the cases where the previous influence events are unordered, such as in the influence sequence $[(P,Q),P]$, and the cases where they are ordered, as in $[Q,P]$.
Recall that a measurement sequence can be decomposed by the series operation $\mydot$ into two subsequences that share a common single influence event as defined in (\ref{eq:series}).  This results in subsequences that consist of a minimum of a pair of influence events.  Here I seek to define a series operator $*$ that can build a sequence by appending a single move, or a \emph{singleton sequence}, to a sequence without regard as to whether the endpoints are influence events or unordered sets of influence events.  Consider appending a $\move{P}$-move to the unordered sequence $[(P,Q)]$ (Figure \ref{fig:diamond-decomposition}A)
\begin{equation}
[(P,Q),P] = [(P,Q)] * [P]
\end{equation}
where the helicity state preceding the final $\move{P}$-move is ambiguous.  To derive a consistent rule for manipulating amplitudes when sequences are joined by the operator $*$, first decompose the $[(P,Q),P]$ using the series and parallel operations so that (Figures \ref{fig:diamond-decomposition}B and C)
\begin{align} \label{eq:ambiguous-initial-state}
[(P,Q),P] &= [P,Q,P] \vee [Q,P,P] \\
&= [Q,P,P] \vee [P,Q,P] \nonumber \\
&= ([Q,P] \mydot [P,P]) \vee ([P,Q] \mydot [Q,P]). \nonumber
\end{align}
I will write the amplitudes assigned to $[Q,P]$ and $[P,Q]$ from the unordered pair, which represent the `initial' helicity states, as $\phi_P$ and $\phi_Q$, respectively, so that I can later consider a wider variety of `initial' states.  In general, I will write the amplitudes assigned to influence sequences $\seq{P}{P}, \seq{P}{Q}, \seq{Q}{P}$, and $\seq{Q}{Q}$ using similar notation, as $[PP], [PQ], [QP]$, and $[QQ]$, respectively, so that each amplitude is easily compared to the sequence to which it corresponds.  The result is that the amplitude associated with the sequence $[(P,Q),P]$ is given by
\begin{equation} \label{eq:amplitude-assignment}
[PP] \phi_P + [QP] \phi_Q
\end{equation}
where on the right-hand side I simply switched the order of the two complex products.
Since for any appended singleton sequence the resulting expression is linear, the new decomposition rule $[(P,Q),P] = [(P,Q)] * [P]$ can be written as a matrix equation where the $\move{P}$- and $\move{Q}$-moves are each associated with a matrix operator, or propagator, with amplitude-valued entries
\begin{equation}
P = \begin{pmatrix} [PP] & [QP] \\ 0 & 0 \end{pmatrix} \qquad \mbox{and} \qquad Q = \begin{pmatrix} 0 & 0 \\ [PQ] & [QQ] \end{pmatrix},
\end{equation}
and the two amplitudes $\phi_P$ and $\phi_Q$ associated with the `initial' sequence are written as a column-vector $\phi$
\begin{equation}
\phi = \begin{pmatrix} \phi_P \\ \phi_Q \end{pmatrix}.
\end{equation}
The amplitude associated with $[(P,Q),P] = [(P,Q)] * [P]$ is then given by
\begin{equation}
P\phi = \begin{pmatrix} [PP] & [QP] \\ 0 & 0 \end{pmatrix} \begin{pmatrix} \phi_P \\ \phi_Q \end{pmatrix} = \begin{pmatrix} [PP]\phi_P + [QP]\phi_Q \\ 0\end{pmatrix},
\end{equation}
which is consistent with (\ref{eq:amplitude-assignment}) obtained by employing the series and parallel operators.

To apply the propagator matrices to a particular sequence, simply employ the parallel operator to expand all unordered sequences and write each resulting ordered sequence in reverse order. Each sequence element is then replaced with the corresponding propagator matrix so that adjacent matrices are to be multiplied.  Last, the parallel operators are replaced by matrix addition operators.  For example, the sequence
$$[(P,Q),P] = [P,Q,P] \vee [Q,P,P]$$
is associated with the propagator given by $(PQP + PPQ)$, which results in the column vector of amplitudes $(PQP+PPQ)\phi$.

The result is that to keep track of the propagation of amplitudes along the influence sequence, one can employ a two-component column vector consisting of the amplitudes associated with each of the two helicity states of the particle, which indicate whether the previous influence event was $\move{P}$ or $\move{Q}$.  In this sense, it is analogous to a \emph{\textbf{Pauli spinor}}.  Extending this to accommodate a particle that both influences and is influenced requires keeping track of four amplitudes, which is suggestive of a Dirac spinor.

\subsection{The Feynman Checkerboard}
I now consider a textbook example, Problem 2.6 on pp. 35-36 of Feynman and Hibbs (1965)\cite{Feynman&Hibbs}, where the ultimate aim is to compute the probability of a free particle of mass $M$ evolving from an initial state $a$ representing the spacetime coordinates $(t_a, x_a)$ to a final state $b$ representing the spacetime coordinates ($t_b, x_b)$.  This is widely known as the Feynman checkerboard problem:

\begin{quoting}
Suppose a particle moving in one dimension can only go forward and backward at the velocity of light.
... 
Then in the $xt$ plane all trajectories shuttle back-and-forth with slopes of $\pm45^{\circ}$...  The amplitude for such a path can be defined as follows: Suppose time is divided into small equal steps of length $\epsilon$. Suppose reversals of direction can only occur at the boundaries of these steps, i.e., at $t = t_a + n\epsilon$, where $n$ is an integer. ... The correct definition [of the amplitude] in the present case is $\phi = (i \epsilon)^R$ ... where $R$ is the number of reversals, or corners, along the path.  ... The definition given here of the amplitude, and the resulting kernel, is correct for a relativistic free theory of a free particle moving in one dimension.  The result is the Dirac equation for that case.\cite[pp. 35-36]{Feynman&Hibbs}
\end{quoting}

As discussed above, the influence-based free particle model reproduces this behavior precisely.  Unordered sequences can be decomposed into a set of all possible ordered sequences, which represent all possible paths that a particle could take from $a$ to $b$.  For each ordered sequence of \move{P} and \move{Q} influence events, the observers interpret the particle as zig-zagging back-and-forth at the maximum invariant speed $\beta = \pm1$ where reversals occur whenever the sequence changes from \move{P} to \move{Q} or vice versa.  Feynman and Hibbs note that the ``correct definition'' of the amplitude should introduce a factor of $i$ for every reversal.  Many authors have demonstrated that this amplitude assignment, and the resulting kernel, leads to the Dirac equation in 1+1 dimensions \cite{Feynman&Hibbs}\cite{gersch1981feynman}\cite{Jacobson+Schulman:1984}\cite{Jacobson:1985}\cite{ord1993dirac}\cite{Kauffman:1996:DiscreteDirac}\cite{ord2002feynman}\cite{Earle:DiracMaster2011}.  Rather than reviewing these results, I will demonstrate that the amplitude assigned by Feynman and Hibbs is readily derived.

%
%
%


\subsection{Deriving Amplitudes}
\begin{figure}[t]
\begin{center}
  \includegraphics[height=0.15\textheight]{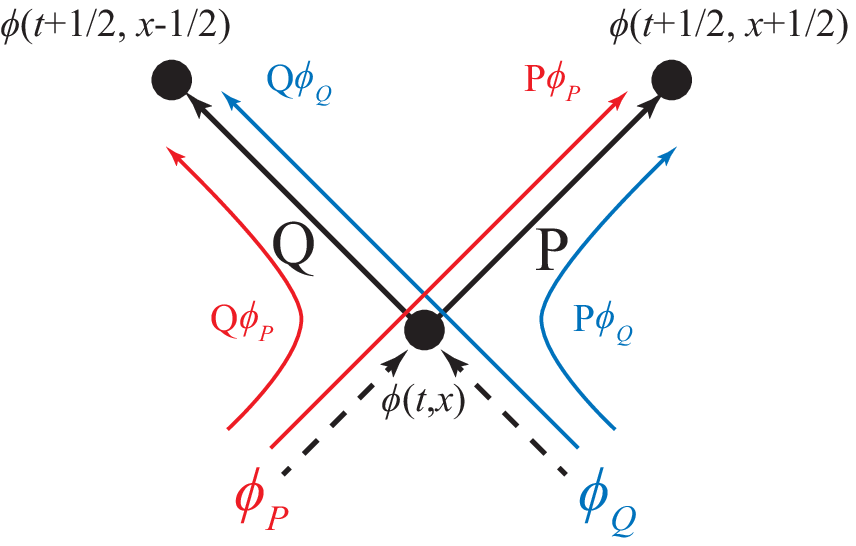}
\end{center}
  \caption{This figure illustrates the four amplitudes that go into relating $\phi(t,x)$ to either $\phi(t+\frac{1}{2}, x-\frac{1}{2})$ or $\phi(t+\frac{1}{2}, x+\frac{1}{2})$, which occurs with probability one.  There are two possible ways in which the initial state can occur with the particle previously having undergone either a \move{P-move} or a \move{Q-move}.  These two initial states are quantified by a pair of amplitudes $\phi_P$ and $\phi_Q$, which are then each propagated by either a $P$ or $Q$ matrix.}
  \label{fig:two-component-spinor-p1}
\end{figure}

The influence-based evolution of a free particle can be understood by considering the transition from one of two possible influence events, \move{P} or \move{Q}, to one of two possible influence events, \move{P} or \move{Q}, as illustrated in Figure \ref{fig:two-component-spinor-p1}.  This is represented in the spacetime picture as a particle experiencing a transition from some initial state given by $(t,x)$ to one of two possible final states: $(t+\frac{1}{2},x-\frac{1}{2})$ or $(t+\frac{1}{2},x+\frac{1}{2})$, which must occur with probability one:
\begin{equation}  \label{eq:prob-one-transition}
Pr\Big(\Big(t+\frac{1}{2},x-\frac{1}{2}\Big) | (t,x) \Big) + Pr\Big(\Big(t+\frac{1}{2},x+\frac{1}{2}\Big) | (t,x) \Big) = 1.
\end{equation}

The probability can be calculated via the Born rule so that (\ref{eq:prob-one-transition}) can be written compactly as
\begin{equation}
(Q \phi)^\dagger (Q \phi) + (P \phi)^\dagger (P \phi) = 1,
\end{equation}
which after simplifying gives
\begin{equation}
\phi^\dagger (Q^\dagger Q + P^\dagger P) \phi = 1.
\end{equation}
This should hold for arbitrary $\phi$, which allows for different probabilities for the particle to arrive from the left or the right, subject to the constraint that these probabilities sum to unity: $\phi^\dagger \phi = 1$.
This implies that
\begin{equation} \label{eq:PQ-condition}
Q^\dagger Q + P^\dagger P = I
\end{equation}
where $I$ is the identity matrix.

I will write the unknown amplitudes $[PP], [QP], [PQ]$ and $[QQ]$ in the propagators $P$ and $Q$ as $x, y, w$ and $z$ respectively, so that
\begin{equation}
P = \begin{pmatrix} x&y \\ 0&0 \end{pmatrix} \qquad \mbox{and} \qquad Q = \begin{pmatrix} 0&0 \\ w&z \end{pmatrix}.
\end{equation}

The condition in (\ref{eq:PQ-condition}) can be rewritten as
\begin{equation}
\begin{pmatrix}
0 & w^* \\
0 & z^* \end{pmatrix}
\begin{pmatrix}
0 & 0 \\
w & z \end{pmatrix}
+
\begin{pmatrix}
x^* & 0 \\
y^* & 0 \end{pmatrix}
\begin{pmatrix}
x & y \\
0 & 0 \end{pmatrix}
=
\begin{pmatrix}
1 & 0 \\
0 & 1 \end{pmatrix}
\end{equation}
which becomes
\begin{equation}
\begin{pmatrix}
w^*w & w^*z \\
z^*w & z^*z \end{pmatrix}
+
\begin{pmatrix}
x^*x & x^*y \\
y^*x & y^*y \end{pmatrix}
=
\begin{pmatrix}
1 & 0 \\
0 & 1 \end{pmatrix}
\end{equation}
resulting in
\begin{equation} \label{eq:complex-norm-condition}
w^{*} w + x^{*} x = 1,
\qquad
z^{*} z + y^{*} y = 1,
\end{equation}
and
\begin{equation} \label{eq:complex-off-diag-condition}
w^{*} z = -x^{*} y,
\qquad
z^{*} w = -y^{*} x.
\end{equation}

Writing $x = a e^{i\alpha}$, $w = b e^{i\beta}$, $y = c e^{i\gamma}$, and $z = d e^{i\delta}$ the equations above reduce to
\begin{align}
a^2 + b^2 &= 1 \\
c^2 + d^2 &= 1 \\
ac &= bd.
\end{align}
With some algebra, it follows that either $a = d$ and $b = c$, or $b = -c$ and $a = -d$, so that without loss of generality, one can write $x = a e^{i\alpha}$, $w = b e^{i\beta}$, $y = b e^{i\gamma}$, and $z = a e^{i\delta}$.

Substituting the expressions for $x, y, w$ and $z$ into (\ref{eq:complex-off-diag-condition}) and dividing through by $ab$ gives
\begin{equation}
e^{i(\delta-\beta)} = - e^{i(\gamma-\alpha)}.
\end{equation}
There is a great deal of freedom in dealing with the phases.  In the special case where the momentum of the particle is zero, the rates $r_P = r_Q$ and there is a symmetry between projections onto the chains $\P$ and $\Q$ such that $w=y$ and $x=z$.  This implies that the phases are related by $\alpha = \delta$ and $\beta = \gamma$ so that the equation above becomes
\begin{equation}
e^{i(\alpha-\beta)} = - e^{i(\beta-\alpha)},
\end{equation}
where the phases $\alpha$ and $\beta$ differ by $90^{\circ}$, resulting in amplitudes that differ by a factor of $i$.
Without loss of generality, I will choose $\alpha = 0$ so that $x = z = a$, and $\beta = \frac{\pi}{2}$ so that $y = w = bi$, where the constants $a^2$ and $b^2$ are related to the probabilities of the particle influencing chains $\P$ and $\Q$, respectively.  In general, the resulting propagators are
\begin{equation} \label{eq:general-propagator}
P = \begin{pmatrix} a&bi \\ 0&0 \end{pmatrix} \qquad \mbox{and} \qquad Q = \begin{pmatrix} 0&0 \\ bi&a \end{pmatrix},
\end{equation}
where $a^2 + b^2 = 1$.
Again, considering the case where the momentum of the particle is zero, symmetry requires that $a = b = \frac{1}{\sqrt{2}}$ so that the propagators become
\begin{equation}
P = \frac{1}{\sqrt{2}}\begin{pmatrix} 1&i \\ 0&0 \end{pmatrix} \qquad \mbox{and} \qquad Q = \frac{1}{\sqrt{2}}\begin{pmatrix} 0&0 \\ i&1 \end{pmatrix},
\end{equation}
where the resulting amplitudes are those suggested by Feynman and Hibbs so that each reversal (transition from $\P$ to $\Q$ or vice versa) brings with it a factor of $i$.  This is known to result in the Dirac equation for the free particle in 1+1 dimensions\cite{Feynman&Hibbs}\cite{gersch1981feynman}\cite{Jacobson+Schulman:1984}\cite{Jacobson:1985}\cite{ord1993dirac}\cite{Kauffman:1996:DiscreteDirac}\cite{ord2002feynman}\cite{Earle:DiracMaster2011}. The more general propagator obtained in (\ref{eq:general-propagator}) reflects a different state of knowledge about the particle where it is known that it will influence $\P$ and $\Q$ with different probabilities.  It is interesting that the matrix entries must be complex, since this results in interference effects that are considered to be inherently quantum mechanical.  Remarkably, this is derived as a consequence of the requirement to make consistent predictions about the free particle influence patterns.

\section{Discussion} \label{sec:Discussion}
Over the last century, it has become increasingly apparent that information and the observer play a central role in physics.  This has led some to go as far as to suggest that information is central to physics \cite{zurek:1990}\cite{brassard:2005information}, which is summarized by Wheeler's aphorism ``It from Bit'' \cite{wheeler:1990}.  While I do not suggest that information creates a physical reality, it seems reasonable to propose that optimization of the processing of information about the world around us ought to guide our description of physical reality.  This has led some to investigate whether the laws of physics are consistent with the optimal processing of information \cite{Brillouin:1956science}\cite{Jaynes:InfoTheory}\cite{Knuth:infophysics}\cite{Caticha2012:entropic}\cite{Goyal:2012information}\cite{Wootters:2013communicating}, while others have focused on the relation between information and physics \cite{Bekenstein:1973blackhole-entropy}\cite{zurek:1990}\cite{tHooft:1993dimensional}\cite{Susskind:1995}\cite{Bousso:2002holographic}\cite{Pawlowski+etal:2009information-causality}\cite{Schumacher+Westmoreland:2010}\cite{Verlinde:2011:entropic-gravity} in areas traditionally distinct from statistical mechanics where the relation has been understood for some time \cite{Brillouin:1956science}\cite{Jaynes:InfoTheory}\cite{Baierlein:1971}\cite{Tribus+McIrvine:1971sciam}.  The latter studies have revealed that entropy is somehow more than relevant to spacetime geometry.  But one is left wondering, ``The entropy of what?'' and moreover, ``Which entropy?'', since there are many entropies, each depending both on the topic of discourse and the chosen level of description \cite{Tseng+Caticha:Gibbs}.  Such findings hint not only at \emph{what we ought to know}, but also at \emph{how we ought to describe it}, in order to answer the questions that are important to us.

This paper summarizes recent efforts by myself and several colleagues that can be divided into two conceptually different aspects.  The first is the consistent apt quantification of algebraic or order-theoretic symmetries.  These have resulted in a mathematical foundation for probability theory \cite{Knuth&Skilling:2012}, quantum mechanics \cite{GKS:PRA}\cite{GKS:me09}\cite{GK:Symmetry}, and more recently, special relativity \cite{Knuth+Bahreyni:EventPhysics}\cite{Bahreyni:Thesis}.  In each of these problems, symmetries constrain consistent apt quantification resulting in constraint equations that represent formal relations that are often considered to be physical laws \cite{Knuth:laws}\cite{Knuth:measuring}.  Not only is the derivation of laws from symmetry, in this way, clean and compelling, but also it supports the vague, yet nearly universal, notion that physical laws reflect an underlying order in the universe.  What is lacking in these derivations is a sense of scale.  Physical constants, such as the speed of light $c$, which relates the measures of time and space, and Planck's constant $h$, which relates the measures of temporal rates and energy, are arbitrary.  Here they are set to unity, which is often referred to as `natural units'.  The fact is that symmetry and consistency as considered here cannot set such scales.

The second aspect arises from the fact that physics as a theory must be about \emph{something}.  Quantifying symmetries amounts to mathematics.  Applying such theories to models that describe some aspect of the universe is physics.  Here I summarize and expound upon recent work \cite{Knuth:fermions} inspired both by our investigations into special relativity and the physics of events \cite{Knuth+Bahreyni:EventPhysics}\cite{Bahreyni:Thesis}, and by an early model of direct particle-particle interaction by Wheeler and Feynman \cite{Wheeler+Feynman:1945}\cite{Wheeler+Feynman:1949}.  I introduce a model in which particles influence one another, and demonstrate that consistent apt quantification of such a model reproduces many familiar particle `properties', such as: position, speed, mass, energy, momentum and helicity.  By conceiving of these `properties' as \emph{descriptions} about how a particle influences others, rather than \emph{attributes} possessed by a particle, it appears that at least some conceptual mysteries related to quantum complementarity, which arises here from the fact that influence can be described in terms of both intervals and rates, and quantum contextuality can be resolved.  Furthermore, I show that the problem concerning inferences made by observers about the influence patterns of a free particle is identical to the Feynman checkerboard model of a fermion \cite{Feynman&Hibbs}.  In this context, the Dirac equation represents a prescription for making optimal inferences about a particle's behavior.

There are several specific assumptions that lead to the Dirac equation in 1+1 dimensional flat spacetime.  It is assumed that there are two coordinated observers.  These observers collectively define a 1+1 dimensional subspace in the poset, which when consistently quantified results in the Minkowski metric.  In addition, we have assumed that the observers remain coordinated when observing a free particle.  However, influence can destroy coordination by changing the momentum and energy of an observer.  This appears to introduce a discrete form of curvature, though it remains to be seen whether such curvature is consistent with general relativity.  To properly assess this, one must consider the problem in 3+1 dimensions.  However, the question as to why the poset itself, or perhaps instead, a consistent description of a poset, is limited to 3+1 dimensions remains open \cite{Knuth+Bahreyni:EventPhysics}.  In addition, the extension of the Feynman checkerboard model to 3+1 dimensions has posed difficulties despite some potential progress \cite{Jacobson:1985}\cite{ord1993dirac}\cite{Bialynicki-Birula:1994}\cite{Earle:DiracMaster2011}.  Given that consistent quantification of the poset of influence events leads to fundamental concepts of geometry and spacetime physics \cite{Knuth+Bahreyni:EventPhysics}, it is perhaps not so surprising that the Dirac equation naturally emerges, since it has been known for some time that there is an intimate relation between spacetime geometry and the Dirac equation \cite{Hestenes:1990zitterbewegung}.

The problem with having to assume a model is that it is, by its very nature, provisional.  As with any theory, it will take further investigation to determine whether this influence-based model of particle-particle interaction is sufficient to describe the wide array of physics that awaits an encompassing foundational theory. Here, at least, I hope to have demonstrated that it may be \emph{possible} to conceive of a simple model that can capture a wide array of physical phenomena, many of which have defied explanation for so long that some have begun to wonder whether or not physics at this level suffers from a fundamental incompleteness \cite{Buchanan:2007incompleteness}.

This work provides a new perspective of Wheeler's ``It from Bit''.  There is an underlying physics---perhaps the physics of influence---that results in observers recording (or being affected by) acts of influence, which correspond to bits of information.  Since these influence events affect different observers, they cannot be ordered, which requires the use of quantum mechanics to compute  probabilities.  One must assign an amplitude to the unordered sequence of influence events, which can be done by considering the relation of the unordered sequence to all possible ordered sequences (equivalently spacetime paths) that can be generated from it.  This has been interpreted as the particle having taken all possible spacetime paths, whereas it is more accurate to recognize that quantum mechanics makes no statement at all as to what the particle does when it is not measured (when influence sequences are unordered) \cite{GK:Symmetry}.  Based on these BITs that originate from influence events (IT), observers make inferences about their model of the universe (IT*), so that $\mbox{IT} \rightarrow \mbox{BIT} \rightarrow \mbox{IT*}$.  Thus information about the influence events is central both in that it is precisely this information that is being described as well as being used to make inferences.  The result is that laws derive from both consistent descriptions and optimal information-based inferences made by embedded observers.

\section*{Acknowledgements}
The author would like to thank Keith Earle, Newshaw Bahreyni, Ariel Caticha, Seth Chaiken, Oleg Lunin, Jeffrey Scargle, John Skilling, Michael Way, and David Wolpert for many insightful discussions and comments.  The author would especially like to thank James Lyons Walsh and the three anonymous reviewers for their valuable comments.  This work was supported by a grant from the Templeton Foundation.

\section*{Notes on contributor}
Kevin H. Knuth is an Associate Professor in the Departments of Physics and Informatics at the University at Albany. He is Editor-in-Chief of the journal Entropy, the co-founder and President of a robotics company, Autonomous Exploration Inc., and a former NASA research scientist. He has more than 15 years of experience in designing Bayesian and maximum entropy-based machine learning algorithms for data analysis applied to the physical sciences.  His interests in the foundations of inference and inquiry has led his theoretical investigations to focus on the foundations of physics from an information-based perspective.

\bibliographystyle{tCPH}
\bibliography{knuth}

\end{document}